\documentclass[12pt]{article}
\usepackage{latexsym}
\usepackage{epsfig,amssymb,euscript,slashed}
\usepackage[linktocpage]{hyperref}
\usepackage{amsmath}
\usepackage[noadjust]{cite}
\usepackage{mathtools}
\usepackage{physics}
\usepackage{extarrows}
\usepackage{xcolor}
\usepackage{float}
\usepackage{graphicx}
\usepackage{subcaption}
\usepackage{tcolorbox}
\usepackage{relsize}
\usepackage{environ}
\usepackage{tikz}
\usepackage[normalem]{ulem}

\tikzset{Witten diagram/.style={execute at begin picture={%
\draw[black] circle[radius=\pgfkeysvalueof{/tikz/Witten/radius}];
}},vertex/.style={circle,fill,inner sep=1.3pt,node
contents={}},box/.style={rectangle,fill,inner sep=2pt,node
contents={}},
Witten/.cd,radius/.initial=1.9cm}
\NewEnviron{wittendiagram}[1][]{\vcenter{\hbox{\begin{tikzpicture}[Witten diagram,#1]%
\BODY
\end{tikzpicture}}}}
\usepackage{array,calc,epsfig}
\usepackage{bbm}
\usepackage{fancybox}

\newcommand{\hb}{\bar{h}}
\newcommand{\pd}{\partial}
\newcommand{\pdb}{\bar{\partial}}

\newcommand{\zb}{\bar{z}}
\newcommand{\Zb}{\bar{Z}}
\newcommand{\jb}{\bar{j}}

\def\dDisc{{\rm dDisc}\,}

\newcommand{\gro}[1]{:\! #1 \!:}

\numberwithin{equation}{section} 
\usepackage[]{graphicx}
\usepackage{color}

\newcommand{\cO}{\mathcal{O}}
\newcommand{\plog}[2]{\mathrm{Li}_{#1} \left( #2 \right) } 
\newcommand{\tn}{(|\ell|+2k)}
\newcommand{\sn}{\ell}
\newcommand{\cI}{\mathcal{I}}
\newcommand{\cJ}{\mathcal{J}}

\newcommand{\cG}{\mathcal{G}}

\def\nb{{\bar{n}}}

\allowdisplaybreaks
\begin{document}
\font\cmss=cmss10 \font\cmsss=cmss10 at 7pt

\begin{flushright}{  
\scriptsize QMUL-PH-21-24}
\end{flushright}
\hfill
\vspace{18pt}
\begin{center}
{\Large 
\textbf{Holographic correlators with multi-particle states
}}

\end{center}

\vspace{8pt}
\begin{center}
{\textsl{Nejc \v{C}eplak$^{\,a}$, Stefano Giusto$^{\,b, c}$, Marcel R. R. Hughes$^{\,d}$ and Rodolfo Russo$^{\,d}$}}

\vspace{1cm}

\textit{\small ${}^a$  Institut de Physique Th\'eorique,
Universit\'e Paris Saclay,\\
CEA, CNRS, F-91191 Gif sur Yvette, France} \\  \vspace{6pt}

\textit{\small ${}^b$ Dipartimento di Fisica,  Universit\`a di Genova, Via Dodecaneso 33, 16146, Genoa, Italy} \\  \vspace{6pt}

\textit{\small ${}^c$ I.N.F.N. Sezione di Genova,
Via Dodecaneso 33, 16146, Genoa, Italy}\\
\vspace{6pt}

\textit{\small ${}^d$ Centre for Research in String Theory, School of Physics and Astronomy\\
Queen Mary University of London,
Mile End Road, London, E1 4NS,
United Kingdom}\\
\vspace{6pt}

\end{center}

\vspace{12pt}

\begin{center}
\textbf{Abstract}
\end{center}

\vspace{4pt} {\small
\noindent 
We derive the connected tree-level part of 4-point holographic correlators in \mbox{AdS$_3\times S^3\times \mathcal{M}$} (where ${\cal M}$ is $T^4$ or $K3$) involving two multi-trace and two single-trace operators. These connected correlators are obtained by studying a heavy-heavy-light-light correlation function in the formal limit where the heavy operators become light. These results provide a window into higher-point holographic correlators of single-particle operators. We find that the correlators involving multi-trace operators are compactly written in terms of Bloch-Wigner-Ramakrishnan functions -- particular linear combinations of higher-order polylogarithm functions. Several consistency checks of the derived expressions are performed in various OPE channels. We also extract the anomalous dimensions and 3-point couplings of the non-BPS double-trace operators of lowest twist at order $1/c$ and find some positive anomalous dimensions at spin zero and two in the K3 case.
}

\vspace{1cm}

\thispagestyle{empty}

\vfill
\vskip 5.mm
\hrule width 5.cm
\vskip 2.mm
{
\noindent {\scriptsize e-mails: {\tt nejc.ceplak@ipht.fr, stefano.giusto@pd.infn.it, m.r.r.hughes@qmul.ac.uk, r.russo@qmul.ac.uk} }
}

\setcounter{footnote}{0}
\setcounter{page}{0}

\newpage

\tableofcontents


\section{Introduction}
\label{sec:intro}

Holographic dualities provide a powerful tool with which to study correlators in strongly coupled conformal field theories (CFTs) in the limit where the number of degrees of freedom becomes large. In this regime -- usually called the large $N$ limit -- it is possible to separate primary operators into a class of ``light'' states, whose conformal dimensions do not grow as $N$ becomes large, and various types of ``heavy'' states, whose conformal weights scale with some power of $N$. Another characterisation of the operator spectrum, relevant for holographic theories at large $N$, is the distinction between single- and multi-particle states. As indicated by their names, this distinction is more readily understood in the dual gravitational description where operators of the first type are dual to single-particle bulk states, while the second class of operators is dual to composite objects of elementary bulk fields. In the best studied example of holographic dualities -- that of ${\cal N}=4$ super Yang-Mills (SYM) and AdS$_5 \times S^5$ -- the CFT description of single-particle states corresponds to single-trace composite operators (up to $1/N$ corrections) and multi-particle states correspond to multi-trace operators. Due to this, we will often use the same nomenclature also in the case of the AdS$_3$/CFT$_2$ holographic duality which will be the focus of this work.

Ever since the early days of their study, the approach of Witten diagrams~\cite{Witten:1998qj} provided an explicit avenue for the calculation of holographic correlators among light single-trace operators and several results were obtained for 4-point correlators in the important example of ${\cal N}=4$ SYM theory~\cite{Liu:1998ty,D'Hoker:1999ea, D'Hoker:1999jc, D'Hoker:1999jp, D'Hoker:1999pj, DHoker:1999mqo}. More recently, knowledge of this class of correlators has been substantially expanded by using a variety of new ideas including: the Mellin space representation~\cite{Penedones:2010ue, Fitzpatrick:2011ia}, the position-space approach introduced in~\cite{Rastelli:2016nze, Rastelli:2017udc,Rastelli:2019gtj}, the use of the large spin expansion~\cite{Alday:2016njk} and the Lorentzian inversion formula of~\cite{Caron-Huot:2017vep,Alday:2017vkk}. Here we will expand on the approach based on ``microstate geometries''~\cite{Galliani:2017jlg,Bombini:2017sge}, which was used in~\cite{Giusto:2018ovt,Giusto:2019pxc,Giusto:2020neo} to derive holographic 4-point correlators between singe-particle states in the case of AdS$_3$/CFT$_2$. The basic idea of this approach is to start from a correlator where a pair of conjugate single-trace operators is made heavy by considering their multi-trace versions obtained by taking a large number of identical elementary constituents. It is then possible to exploit several known asymptotically AdS$_3 \times S^3$ smooth solutions of type IIB supergravity~\cite{Lunin:2001fv, Kanitscheider:2007wq,Bena:2015bea, Bena:2017xbt} dual to this class of heavy states. In particular, from quadratic fluctuations around these supergravity solutions it is possible -- by using the standard AdS/CFT dictionary -- to derive so-called HHLL 4-point correlators with two heavy states corresponding to the geometry and two light states corresponding to the fluctuations. The geometries describing these microstates depend on a set of parameters which quantify the number and type of constituents forming the heavy states. An observation of~\cite{Giusto:2018ovt,Giusto:2019pxc,Giusto:2020neo} is that it is possible to take a ``light'' limit by formally setting the number of constituents of the heavy states to one and -- despite bringing the smooth solutions outside the regime of validity of supergravity -- the limit is smooth at the level of correlators and produces results that have all the expected features of 4-point AdS$_3$/CFT$_2$ correlators among single-particle states. Some of these correlators were calculated in~\cite{Rastelli:2019gtj} by different techniques, providing an independent cross-check of the approach discussed above.

In this paper we extend the analysis of the light limit of HHLL correlators beyond the first non-trivial order in the number of constituents and show that the subleading corrections capture interesting information about 4-point correlators that involve two light, but multi-trace operators, and two standard single-trace operators. As is usual for holographic results obtained with the use of a supergravity approximation, these correlators are valid in the large $N$, strong coupling regime of the CFT. To be more precise, we provide evidence that the results obtained from the light limit of the supergravity HHLL correlators capture the {\em tree-level connected Witten diagram contributions} relevant for the correlators under study. It is interesting to see that the explicit results involve a generalisation of the $D$-function usually appearing in the correlators among single-particle operators. These $D$-functions arise naturally from the integrals appearing in a Witten diagram describing the contact 4-point interaction between bulk fields and can be written in terms of the Bloch-Wigner dilogarithm~\cite{zagier1990bloch}. The multi-trace correlators we obtain here are written in terms of the Bloch-Wigner-Ramakrishnan polylogarithms~\cite{zagier1991polylogarithms} which are generalisations sharing several properties of the standard Bloch-Wigner function, but involve higher order polylogarithms. These polylogarithms have also appeared in other physics applications, such as the evaluation of multi-loop Feynman integrals~\cite{Usyukina:1993ch}, the analysis of the free energies in $O(N)$ and $U(N)$ models~\cite{Sachdev:1993pr,Filothodoros:2016txa} and, in the holographic context, the expression of integrated four-point correlators~\cite{Dorigoni:2021guq}.

As already mentioned above, the multi-trace operators we consider are made from identical constituents which are mutually BPS and so it is possible to relate the 4-point correlators we discuss to a particular kinematic limit of a higher-point function involving just single-particle states. This relation to higher-point functions involves two steps. The first is the kinematic limit taking two groups of $n$ identical operators in a $(2n+2)$-point correlator to the same position and the second is to relate the resulting $n$-trace operators to those that are natural from the point of view of the heavy operators dual to the smooth bulk geometries which are used in deriving the HHLL correlators. Despite this second step, we argue that the functions appearing in the multi-trace correlators we construct are present also in higher-point functions of single-trace operators. As will be discussed in more detail in the main text, this picture makes it evident that these correlators contain both classical (tree-level) and quantum (loop) contributions in the gravity picture at a given order in the $1/N$-expansion. However, as previously stressed, our results capture only the classical part and provide a window on the structure of tree-level Witten diagrams for correlators with six or more single particle external states, albeit in the simplifying kinematic limit where only two cross-ratios survive. We point out that, even in this simplified regime, these results are qualitatively different from those obtained in~\cite{Goncalves:2019znr} where an explicit five-point correlator was calculated in AdS$_5 \times S^5$ and the result could still be written in terms of standard Bloch-Wigner functions.

Since the extrapolation of the HHLL correlator to small values of the heavy operator's conformal dimension -- on which we base our derivation of the correlators with multi-particle states -- is a priori unjustified, it is important to gather independent evidence on the correctness of our conjecture. We thus take various OPE limits of the multi-particle correlators and verify that we obtain consistent results. One can, for example, focus on the light-cone OPE limit -- where to leading order, only the conserved currents are exchanged -- and check that our correlators reproduce the expected behaviour of their conformal blocks. Alternatively, one can isolate the exchange of protected multi-trace operators -- produced in the OPE of operators preserving the same supersymmetries -- and match the corresponding three-point couplings with those computed in the weakly coupled (orbifold) CFT. New dynamical information is contained in the anomalous dimensions and couplings of the non-BPS multi-trace operators and despite, at present, not knowing all of the necessary correlators to extract this information completely, we verify some qualitative features of the OPE in the non-protected channels and derive constraints on these dynamical quantities. We note, however, that the OPE data of the double-trace non-BPS operators with minimal (bare) twist can be inferred from known amplitudes of four single-trace chiral primary operators with the lowest dimension $(1/2,1/2)$ -- these sit in one of the $N_f$ tensor multiplets (where $N_f=5$ or $21$ for the theory compactified on $T^4$ or $K3$). Unlike in AdS$_5\times S^5$ \cite{Aprile:2017xsp}, this task already involves a non-trivial mixing problem\footnote{In a recent article \cite{Aprile:2021mvq} the nice observation was made that, for non-BPS multi-traces in non-trivial flavour representations, the same mixing problem can be solved for arbitrary twist using the available correlator data.}
between the different multiplet flavours. This analysis provides a somewhat surprising result: we find a {\it positive} anomalous dimension for operators of spin 0 {\it and 2}. The question of the existence of a consistent large $N$ CFT with a spin-two operator with a positive anomalous dimension was raised in \cite{Alday:2017gde} and this is, to the best of our knowledge, the first affirmative answer to this search. 

The paper is structured as follows. Section~\ref{sec:bg} introduces the multi-trace $4$-point functions that we will be concerned with in this paper and presents the general idea of their extraction from supergravity HHLL correlators computed using asymptotically AdS$_3 \times S^3$ supergravity solutions. In Section~\ref{sec:mtc} we detail explicit correlators involving two $n$-trace operators for low values of $n$ and explain the connection to correlators involving multi-trace operators with an interpretation as a particular limit of higher-point functions. We also collect the main definitions and properties of the Bloch-Wigner-Ramakrishnan functions in terms of which these correlation functions are naturally written. This is then followed in Section~\ref{sec:lchecks} by an analysis of various kinematic limits of the $n=2$ correlator to demonstrate that they yield the behaviour expected from its identification with the connected correlator between two single-trace and two double-trace operators. We analyse both protected and non-protected quantities. Focusing instead on the standard 4-point functions with single-traces ($n=1$), we show that the full mixing problem in flavour space can be solved and we find the anomalous dimensions for the lowest twist non-BPS double-trace operators in any flavour representation. We note that flavour-singlet operators of spin 0 and 2 have positive anomalous dimensions for the theory compactified on $K3$. Finally, Section~\ref{sec:conclusions} contains a brief summary and an outlook. Appendix~\ref{app:corder} provides details of the resummation process from which our correlators are derived, Appendix~\ref{app:n1summ} gives the explicit form of $n=1$ correlators used in the unmixing of Section~\ref{sssec:n1} and Appendix~\ref{app:inversion} summarises the derivation of the double-trace OPE data from the inversion formula.

\section{Correlators with multi-trace operators: the setup}
\label{sec:bg}

\begin{figure}[t]
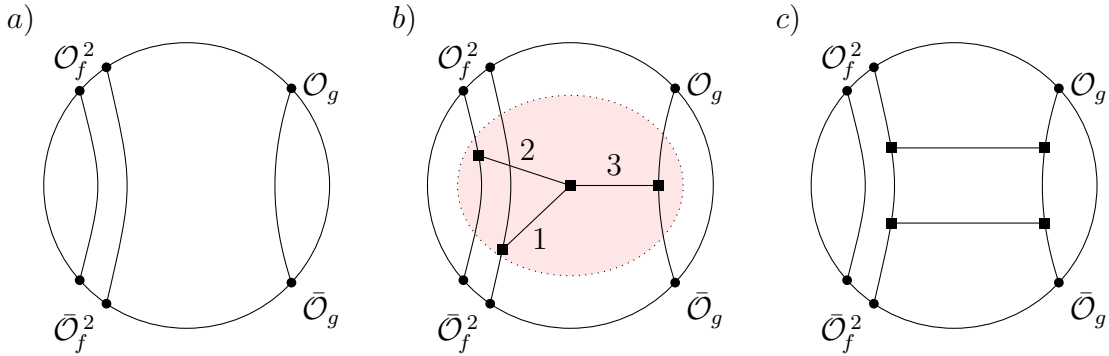

\begin{center}
\[\begin{wittendiagram}
    \draw (1.3928,1.29233) node[vertex] node[anchor=west]{$\mathcal{O}_g$} .. controls (1.1,0.4) and (1.1,-0.4) .. (1.3928,-1.29233) node[vertex]node[anchor=north west]{$\bar{\mathcal{O}}_g$};
    \draw (-1.42217,-1.25993) node[vertex] .. controls (-1.1,0) .. (-1.42217,1.25993) node[vertex]node[anchor=south]{$\mathcal{O}^{\,2}_{\!f}\ $};
    \draw (-1.06626,-1.57261) node[vertex]node[anchor=north east]{$\bar{\mathcal{O}}^{\,2}_{\!f}$} .. controls (-0.7,0) .. (-1.06626,1.57261) node[vertex];
    \node[]  at (-2.2,2.2) {$a)$};
\end{wittendiagram}
\quad
\begin{wittendiagram}
    \fill[red!10] (0,0) ellipse (1.5 and 1.2);
    \draw[dotted,black] (0,0) ellipse (1.5 and 1.2);
    \draw (1.3928,1.29233) node[vertex] node[anchor=west]{$\mathcal{O}_g$} .. controls (1.1,0.4) and (1.1,-0.4) .. (1.3928,-1.29233) node[vertex]node[anchor=north west]{$\bar{\mathcal{O}}_g$};
    \draw (-1.42217,-1.25993) node[vertex] .. controls (-1.1,0) .. (-1.42217,1.25993) node[vertex]node[anchor=south]{$\mathcal{O}^{\,2}_{\!f}\ $};
    \draw (-1.06626,-1.57261) node[vertex]node[anchor=north east]{$\bar{\mathcal{O}}^{\,2}_{\!f}$} .. controls (-0.7,0) .. (-1.06626,1.57261) node[vertex];
    \draw (0,0)node[box] -- node[midway,above]{3} (1.165,0) node[box];
    \draw (0,0) -- node[midway,below]{\:1} (-0.9,-0.85) node[box];
    \draw (0,0) -- node[midway,above]{\:2} (-1.22,0.4) node[box];
    \node[]  at (-2.2,2.2) {$b)$};
\end{wittendiagram}
\quad
\begin{wittendiagram}
    \draw (1.3928,1.29233) node[vertex] node[anchor=west]{$\mathcal{O}_g$} .. controls (1.1,0.4) and (1.1,-0.4) .. (1.3928,-1.29233) node[vertex]node[anchor=north west]{$\bar{\mathcal{O}}_g$};
    \draw (-1.42217,-1.25993) node[vertex] .. controls (-1.1,0) .. (-1.42217,1.25993) node[vertex]node[anchor=south]{$\mathcal{O}^{\,2}_{\!f}\ $};
    \draw (-1.06626,-1.57261) node[vertex]node[anchor=north east]{$\bar{\mathcal{O}}^{\,2}_{\!f}$} .. controls (-0.7,0) .. (-1.06626,1.57261) node[vertex];
    \draw (-0.83,0.5)node[box] -- (1.2,0.5) node[box];
    \draw (-0.83,-0.5)node[box] -- (1.2,-0.5) node[box];
    \node[]  at (-2.2,2.2) {$c)$};
\end{wittendiagram}\]
\end{center}
\caption{Three Witten diagrams contributing to~\eqref{eq:multiTcorr} for the case of $n=2$. 
Diagram $a)$ is disconnected and thus contributes at leading order in large $N$, while the remaining two involve four bulk vertices (denoted by a black square) and so are both suppressed by a factor of $1/N^2$ with respect to the first.
Diagram $b)$ depicts a connected tree-level diagram, while $c)$ contains a disconnected 1-loop structure.
The results derived in this paper focus on the former class of contributions, where in the shaded region one considers all possible ways of obtaining a connected tree-level structure.
Such diagrams contain up to three bulk-to-bulk propagators (labelled explicitly by $1,2,3$ in $b)$) and, as mentioned in the text, this makes a direct evaluation of such contributions challenging. \label{Fig:WDB4}}
\end{figure}

The main object of study in this paper is a special class of holographic correlators in the CFT$_2$ dual to type IIB superstring theory on AdS$_3 \times S^3 \times {\cal M}$, where ${\cal M}$ can be either $T^4$ or $K3$. A first way to characterise this class of correlators is in terms of 4-point functions involving two BPS-conjugate multi-trace operators and two BPS-conjugate single-trace operators. Since all operators that we consider are scalars, conformal invariance implies that the correlators depend on a single function of two cross-ratios
\begin{equation} \label{eq:multiTcorr}
    \langle \bar{\mathcal{O}}^{\,n}_{\!f}(z_1,\bar{z}_1)\,\mathcal{O}^{\,n}_{\!f}(z_2,\bar{z}_2)\,\mathcal{O}_g(z_3,\bar{z}_3) \, \bar{\mathcal{O}}_g(z_4,\bar{z}_4) \rangle = \frac{{\cal G}_{n}(z,\bar{z})}{|z_{12}|^{2n \Delta_f}|z_{34}|^{2 \Delta_g}}\;,
\end{equation}
where $z_{ij}=z_i-z_j$ and $\mathcal{O}_{\!f,g}$ ($\bar{\mathcal{O}}_{\!f,g}$) are (anti)-chiral Primaries Operators (CPO) in the D1-D5 CFT$_2$ with identical holomorphic and antiholomorphic dimensions $h=\bar{h}=\Delta/2$ and flavour indices\footnote{Since the 6D bulk theory has 16 supercharges, the fields organise into different multiplets; we focus on CPOs of the matter tensor multiplets of which there are $5$ ($21$) different flavours in the $\mathcal{M}=T^4$ ($K3$) case.} $f,g$. We define the cross-ratio $z$ as follows
\begin{equation}
  \label{eq:crossratio}
  z = \frac{z_{14}z_{23}}{z_{13}z_{24}} \ ,
\end{equation}
and so it is often convenient to work in the gauge
\begin{equation} \label{eq:zgauge}
    z_1 = 0 \;, \quad z_2 \to\infty\;, \quad z_3 = 1 \;, \quad z_4 =z  \;,
\end{equation}
where the correlator takes the form
\begin{equation}
	\begin{aligned}
	\label{eq:defCn}
	\mathcal{C}_n(z, \zb) \equiv \lim_{\substack{z_2\rightarrow\infty}} |z_2|^{2n \Delta_f}
	\langle \bar{\mathcal{O}}^{\,n}_{\!f}(0,0)\,\mathcal{O}^{\,n}_{\!f}(z_2,\bar{z}_2)\,\mathcal{O}_{\!g}(1,1) \, \bar{\mathcal{O}}_{\!g}(z,\bar{z}) \rangle =\frac{{\cal G}_{n}(z,\bar{z})}{|1-z|^{2 \Delta_g}}
	& \;.
	\end{aligned}
\end{equation}
The function ${\cal G}_{n}(z,\bar{z})$, which contains the dynamical information, at least in principle can be calculated in the holographic regime by summing Witten diagrams. As depicted in Fig.~\ref{Fig:WDB4}, there are different types of contributing diagrams and we will focus on the connected tree-level diagrams, such that in $b)$, which are of order $1/N^n$. For the class of correlators in~\eqref{eq:multiTcorr} there are also other contributions relevant at the same order in the $1/N$ expansion (see the disconnected 1-loop diagram $c$ in Fig.~\ref{Fig:WDB4}) and so our results do not in general represent the full holographic correlators~\eqref{eq:multiTcorr} at order $1/N^n$. We will discuss below how these disconnected loop contributions cancel in our approach.

\begin{figure}[t]
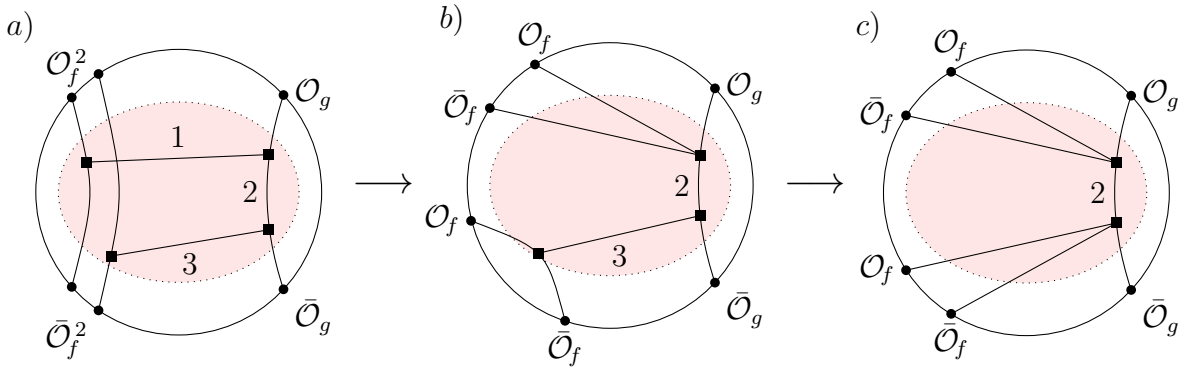

\begin{center}
\[\begin{wittendiagram}
    \fill[red!10] (0,0) ellipse (1.6 and 1.2);
    \draw[dotted,black] (0,0) ellipse (1.6 and 1.2);
    \draw (1.3928,1.29233) node[vertex] node[anchor=west]{$\mathcal{O}_g$} .. controls (1.1,0.4) and (1.1,-0.4) .. (1.3928,-1.29233) node[vertex]node[anchor=north west]{$\bar{\mathcal{O}}_g$};
    \draw (-1.42217,-1.25993) node[vertex] .. controls (-1.1,0) .. (-1.42217,1.25993) node[vertex]node[anchor=south]{$\mathcal{O}^{\,2}_{\!f}\ $};
    \draw (-1.06626,-1.57261) node[vertex]node[anchor=north east]{$\bar{\mathcal{O}}^{\,2}_{\!f}$} .. controls (-0.7,0) .. (-1.06626,1.57261) node[vertex];
    \draw (-0.9,-0.85) node[box] -- node[midway,below]{3} (1.2,-0.5)node[box];
    \draw (-1.22,0.4)node[box] -- node[midway,above]{1} (1.2,0.5)node[box];
    \draw (1.2,0)node[anchor=east]{2};
    \node[]  at (-2.1,2.2) {$a)$};
\end{wittendiagram}
\ \mathlarger{\longrightarrow} 
\begin{wittendiagram}
    \fill[red!10] (0,0) ellipse (1.6 and 1.2);
    \draw[dotted,black] (0,0) ellipse (1.6 and 1.2);
    \draw (1.3928,1.29233) node[vertex] node[anchor=west]{$\mathcal{O}_g$} .. controls (1.1,0.4) and (1.1,-0.4) .. (1.3928,-1.29233) node[vertex]node[anchor=north west]{$\bar{\mathcal{O}}_g$};
    \draw (-1,1.61) node[vertex]node[anchor=south]{$\mathcal{O}_{\!f}$} -- (1.2,0.4)node[box];
    \draw (-1.6,1.03)node[vertex]node[anchor=east]{$\bar{\mathcal{O}}_{\!f}$} -- (1.2,0.4)node[box];
    \draw (-1.85,-0.47) node[vertex]node[anchor=east]{$\mathcal{O}_{\!f}$} .. controls (-0.85,-0.85) .. (-0.6,-1.8) node[vertex]node[anchor=north]{$\bar{\mathcal{O}}_{\!f}$};
    \draw (1.2,0)node[anchor=east]{2};
    \draw (-0.95,-0.9)node[box] -- node[midway,below]{3} (1.2,-0.4)node[box];
    \node[]  at (-2.1,2.2) {$b)$};
\end{wittendiagram}
\ \mathlarger{\longrightarrow}
\begin{wittendiagram}
    \fill[red!10] (0,0) ellipse (1.6 and 1.2);
    \draw[dotted,black] (0,0) ellipse (1.6 and 1.2);
    \draw (1.3928,1.29233) node[vertex] node[anchor=west]{$\mathcal{O}_g$} .. controls (1.1,0.4) and (1.1,-0.4) .. (1.3928,-1.29233) node[vertex]node[anchor=north west]{$\bar{\mathcal{O}}_g$};
    \draw (-1,1.61) node[vertex]node[anchor=south]{$\mathcal{O}_{\!f}$} -- (1.2,0.4)node[box];
    \draw (-1.6,1.03)node[vertex]node[anchor=east]{$\bar{\mathcal{O}}_{\!f}$} -- (1.2,0.4)node[box];
    \draw (1.2,0)node[anchor=east]{2};
    \draw (-1.6,-1.03) node[vertex]node[anchor=east]{$\mathcal{O}_{\!f}$} -- (1.2,-0.4)node[box];
    \draw (-1,-1.61)node[vertex]node[anchor=north]{$\bar{\mathcal{O}}_{\!f}$} -- (1.2,-0.4)node[box];
    \node[]  at (-2.1,2.2) {$c)$};
\end{wittendiagram}
\]
\end{center}
\caption{An example of a connected tree-level Witten diagram contributing to the multi-trace correlator $\mathcal{C}_{n=2}$ with three bulk-bulk propagators which cannot be reduced to a finite sum of contact diagrams using the results of \cite{DHoker:1999mqo}. After two applications of this technique, the remaining bulk-bulk propagator $2$ in the rightmost diagram cannot be reduced further due to being connected to three boundary points at each vertex. After each step given, the result will be proportional to a finite sum of such diagrams.}\label{Fig:2}
\end{figure}

As Fig.~\ref{Fig:WDB4} suggests, it is possible to view the correlators in~\eqref{eq:multiTcorr} as a particular kinematic limit of a $(2n+2)$-point function where the multi-trace operators are replaced by $n$ identical CPOs at different positions, which are then taken to the same point, for instance as $\mathcal{O}^{\,n}_{\!f}(z_2,\bar{z}_2) \sim \lim_{w_a\to z_2} \prod_{a=1}^n \mathcal{O}_{\!f}(w_a,\bar{w}_a)$. Thus our results provide a first window on higher-point holographic correlators\footnote{In~\cite{Green:2020eyj} a particular class of $n$-point functions in $\mathcal{N}=4$ SYM was considered. These correlators violate maximally the $U(1)_Y$ bonus symmetry of the supergravity limit and are related to lower point correlators by a recursion relation. The multi-particle correlators relevant for our analysis preserve the corresponding bonus symmetry, which for the AdS$_3$ case is $SU(2)$, and do not obey any simple recursion relation.}, making it possible to go beyond the explicit 5-point AdS$_5 \times S^5$ example discussed in~\cite{Goncalves:2019znr}, since the $n=2$ case is already related to a correlator involving six single-trace operators. Correlators with six or more external points and their OPE limits discussed in this article are expected to be intrinsically more complicated than the lower point examples since the method introduced in~\cite{DHoker:1999mqo} to deal with bulk-to-bulk propagators will not be sufficient to evaluate them. This can be seen by considering the connected tree-diagram given in Fig.~\ref{Fig:2}a. By following~\cite{DHoker:1999mqo} one can, for instance, write the part of the diagram involving one pair of the operators $\mathcal{O}_{\!f}$, $\bar{\mathcal{O}}_{\!f}$ and the bulk-to-bulk propagator $1$ as a {\em finite} sum of boundary-to-bulk propagators directly linking the two positions of the boundary operators with the bulk interaction vertex (giving the diagram of Fig.~\ref{Fig:2}b). This procedure can be repeated for the remaining pair of the operators $\mathcal{O}_{\!f}$, $\bar{\mathcal{O}}_{\!f}$ and the bulk-to-bulk propagator $3$ to get a sum of diagrams of the form of Fig.~\ref{Fig:2}c. However, after this second step it is not possible to eliminate the final bulk-to-bulk propagator (labelled $2$) because each of its endpoints is connected to three external points. Thus, tree diagrams such as the one depicted in Fig.~\ref{Fig:2}a cannot be recast as a sum of contact diagrams. This is in contrast to the connected tree-diagram given in Fig.~\ref{Fig:WDB4}b, which can be reduced to a finite sum of contact diagrams following the method of \cite{DHoker:1999mqo}. Therefore, even after taking the kinematic limit reducing the higher-point correlators to the $4$-point function~\eqref{eq:multiTcorr}, we do not expect that the contribution of all connected tree diagrams can be expressed in terms of the standard $D_{\Delta_1,\Delta_2,\Delta_3,\Delta_4}$ functions. Indeed as discussed in the next section, new objects (the Bloch-Wigner-Ramakrishnan polylogarithms) naturally appear in this case.

We conclude this introductory section by presenting an effective way to isolate the tree-level connected contributions to~\eqref{eq:multiTcorr} which is suggested by the dual gravitational description for the ``heavy states'' (where $n \sim N$). To have a well-defined semiclassical description it is natural to consider particular coherent-state-like linear combinations of multi-particle states~\cite{Skenderis:2006ah,Kanitscheider:2007wq,Giusto:2015dfa} that are dual to a class of asymptotically AdS$_3 \times S^3$ supergravity solutions. Since these coherent states are classical supergravity objects, one can expect that they receive contributions only from the tree-level diagrams such as the ones in Fig.~\ref{Fig:WDB4}b and \ref{Fig:2}a and that loop diagrams cancel out. We will provide evidence that this is indeed the case. Then by following~\cite{Skenderis:2006ah,Kanitscheider:2007wq,Giusto:2015dfa}, we define the operator
\begin{equation}\label{eq:OH}
    \mathcal{O}_{\!H\!,f} = \sum_{p=0}^N \sqrt{\binom{N}{p}} \left(\frac{B}{\sqrt{N}}\right)^p \left(1-\frac{B^2}{N}\right)^{\frac{N-p}{2}}  \mathcal{O}^{\,p}_{\!f}\;,
\end{equation}
where $B$ is a parameter that, for simplicity, we take real and $\mathcal{O}^{\,p}_{\!f}$ is defined such that the $2$-point function with its conjugate operator is normalised to one. Then we can write an expansion linking the $4$-point HHLL correlator involving the heavy operators $\mathcal{O}_{\!H\!,f}$ in~\eqref{eq:OH} and the correlators $\mathcal{C}_p$ containing the multi-particle constituents $\mathcal{O}^{\,p}_{\!f}$, as follows:
\begin{equation}\label{eq:HHLLseries}
\begin{aligned}
    \langle \bar{\mathcal{O}}_{\!H\!,f} \mathcal{O}_{\!H\!,f} \mathcal{O}_{\!g} \bar{\mathcal{O}}_{\!g} \rangle&=\sum_{p=0}^N \left(\frac{B^2}{N}\right)^p \left(1-\frac{B^2}{N} \right)^{N-p} \binom{N}{p} \,\mathcal{C}_p \\
    &=\sum_{n=0}^N \left(\frac{B^2}{N}\right)^{\!n} \sum_{p=0}^{n} (-1)^{n-p} \,\binom{N-p}{n-p}\,\binom{N}{p} \,\mathcal{C}_p\\
    &=\sum_{n=0}^N \left(\frac{B^2}{N}\right)^{\!n} \binom{N}{n} \,C_{n} \ ,
\end{aligned}
\end{equation}
where
\begin{equation}\label{eq:defconn}
    C_{n} \equiv \sum_{p=0}^{n} (-1)^{n-p} \,\binom{n}{p}\,\mathcal{C}_p\ .
\end{equation}
In the first step we simply expanded the $(N-p)$-th power and rearranged the sums so as to collect the factors of $B^2$ and in the final step we defined the combination $C_{n}$. This is the combination that is encoded in the HHLL correlator and, as mentioned above, it captures only the {\em connected tree-level} diagrams of $\mathcal{C}_n$ for each value of $n$. Let us see how this looks in the simplest examples. For $n=1$ we have
\begin{equation}
  \label{eq:n=1ex}
    C_{1} = \langle \bar{\mathcal{O}}_{\!f}\mathcal{O}_{\!f} \mathcal{O}_{\!g} \bar{\mathcal{O}}_{\!g} \rangle-\langle \mathcal{O}_{\!g} \bar{\mathcal{O}}_{\!g}\rangle\;,
\end{equation}
so the disconnected contribution to the first term is cancelled exactly by the $2$-point function appearing in the second term (note that we have chosen a normalisation such that \mbox{$\langle \bar{\mathcal{O}}_{\!f}(0) \mathcal{O}_{\!f}(\infty)\rangle=1$}). The first example involving multi-trace operators requires $n=2$ and from the definition~\eqref{eq:defconn} we have
\begin{equation}
  \label{eq:n=2ex}
  C_{2} = \langle \bar{\mathcal{O}}_{\!f}^{\,2} \mathcal{O}_{\!f}^{\,2} \mathcal{O}_{\!g} \bar{\mathcal{O}}_{\!g} \rangle -2 \langle \bar{\mathcal{O}}_{\!f} \mathcal{O}_{\!f} \mathcal{O}_{\!g} \bar{\mathcal{O}}_{\!g} \rangle + \langle \mathcal{O}_{\!g} \bar{\mathcal{O}}_{\!g} \rangle\;.
\end{equation}
The fully disconnected contributions of order $N^0$ cancel as before, but now there is also a cancellation among the partially connected contributions -- diagrams with a $2$-point function factorised from the other four operators, which have a leading contribution of order $N^{-1}$ and a 1-loop term of order $N^{-2}$. The remainder of the connected part of this correlator is thus the $N^{-2}$ terms not containing loops.

\section{Multi-trace correlators}
\label{sec:mtc}

One way of calculating a HHLL correlator of the type \eqref{eq:HHLLseries} requires solving a wave equation in the background of the geometry dual to the heavy state. In general this is a difficult problem that cannot be solved exactly and one usually resorts to approximation schemes to simplify the task.
For example, one can use a WKB method~\cite{Bena:2019azk, Bena:2020iyw} or alternatively work in the regime where the heavy state generates only small deformations of AdS$_3\times S^3$ rather than a fully backreacted geometry.
In the latter case, expressions can be obtained analytically using standard holographic techniques and correlators have been obtained employing this method for various heavy states~\cite{Galliani:2016cai, Galliani:2017jlg, Bombini:2019vnc}.
However, such results are typically limited only to first order corrections around the AdS$_3\times S^3$ vacuum and can be related to correlators of single-particle operator~\cite{Giusto:2018ovt,Giusto:2019pxc,Giusto:2020neo}.

Nonetheless, for a specific choice of heavy state it was found that one can evaluate the heavy-heavy-light-light correlator exactly in terms of a double Fourier series \cite{Bombini:2017sge}. 
In this section we study this particular correlator and use the expansion \eqref{eq:HHLLseries} to extract a closed form expression for connected tree-level correlation functions with double-trace operator insertions.
We find that particular combinations of higher-order polylogarithms -- the Bloch-Wigner-Ramakrishnan polylogarithm functions -- appear in these multi-trace correlators and we dedicate a part of this section to a brief review of these functions.

Some of the more technical results, together with examples of correlators involving pairs of higher multi-trace operators can be found in Appendix~\ref{app:corder}.

\subsection{Explicit statement of the higher-order correlator}

To specify precisely which operators appear in the correlator that is to be the main focus of this section, we use the notation of the orbifold point of the D1-D5 CFT at which it is described by a sigma model with target space $\left(\mathcal{M}\right)^N/S_N$ (most of our results are valid for both $\mathcal{M}=T^4$ and $K3$). 
The theory contains a collection of $N$ free bosonic and fermionic fields which naturally split into a holomorphic and an anti-holomorphic sector as 
\begin{align}
	\label{eq:freefieldsdef}
	\left(\partial X_{(r)}^{A \dot A}(z)\,, \psi_{(r)}^{\alpha \dot A}(z)\right)\,, \qquad 
	\left(\bar \partial X_{(r)}^{A \dot A}(\zb)\,, \tilde \psi_{(r)}^{\dot \alpha \dot A}(\zb)\right)\,,
\end{align}
where the the indices $A, \dot A =1,2$ are related to the $SU(2)$ symmetries of the internal manifold $\mathcal{M}$; $\alpha, \dot \alpha = \pm$ are fundamental $SU(2)_L\times SU(2)_R$ $R$-symmetry indices; and $(r)= 1, 2, \ldots, N$ is an $S_N$ index labelling the $N$ copies of the target space. 
In this work we focus on operators in the untwisted sector of the orbifold theory and thus all fields are periodic under $(z,\bar z)\to (z \,e^{2\pi i}, \bar z  \,e^{-2\pi i})$.
For more details, see for example \cite{David:2002wn, Avery:2010qw} or appendix A of \cite{Giusto:2015dfa}, whose conventions we follow. 

The point in the moduli space of the D1-D5 CFT that admits a dual semiclassical supergravity description is an infinite distance away from the free orbifold point (see for example \cite{Larsen:1999uk}). Despite this, the notation of the latter can be useful to describe protected supersymmetric operators such as those appearing in our correlators \cite{Baggio:2012rr}.
With that in mind, we consider light operators that are chiral primaries; since the AdS$_3\times S^3$ theory is not maximally supersymmetric, chiral primaries can originate from different 6D supergravity multiplets and in this paper we restrict to those of the $N_f$ tensor multiplets (where either $N_f=5$ or $21$ for $\mathcal{M}=T^4$ and $K3$ respectively). On top of this, we consider such operators with minimal conformal dimension, i.e. with $\Delta=1$. For example, four of the five $T^4$ tensor multiplet CPOs are given in the free orbifold language by
\begin{align}
	\label{eq:OLexp}
	\mathcal{O}_{\!f}(z,\zb) = \frac{\sigma^{(f)}_{\!\dot A \dot B}}{\sqrt{2N}}\sum_{r=1}^{N}\, \psi_{(r)}^{+ \dot A}(z) \, \tilde \psi_{(r)}^{+ \dot B}(\zb)=\frac{1}{\sqrt{N}}\sum_{r=1}^N \mathcal{O}_{\!f (r)}\,,
\end{align}
where $\sigma^{(f)}_{\!\dot A \dot B}$ form a basis of 2 by 2 matrices (the fifth CPO is given by the twist field of order 2). We use the same operator, but with a different flavour index, as the fundamental building block for the construction of the heavy states. As discussed in the previous section the most natural way to define the multi-particle operators in~\eqref{eq:OH} is to take the OPE limit of the single particle states which is regular as all constituents are mutually BPS. It turns out that this is not the choice appropriate for defining the heavy states dual to the supergravity solutions used in~\cite{Galliani:2017jlg,Bombini:2017sge} to derive the HHLL correlators. It is easier to characterise these multi-particle states in terms of the free orbifold theory, where for instance the relevant double-trace operators are $:\!\mathcal{O}^{\,2}_{\!f}\!:\, \sim \sum_{r< s} \mathcal{O}_{ (r)} \mathcal{O}_{(s)}$. Notice that the contribution $r=s$ is absent, while a term of this type arises when taking the OPE limit $\mathcal{O}^{\,2}\, \sim \lim_{w\to z} \mathcal{O}(w) \mathcal{O}(z)$. By following the dictionary~\cite{Taylor:2007hs,Rawash:2021pik} for supersymmetric operators between the supergravity and the free orbifold points, we have the following schematic form for the relation between the two types of double-trace operators discussed above
\begin{equation}
  \label{eq:O:O:}
  \gro{\cO^2} \;= \cO^2 + \frac{1}{\sqrt{N}}\Big[\mbox{single-particle}\Big] +  \frac{1}{N} \Big[\mbox{double-particle}\Big]\;,
\end{equation}
where the operators appearing on the right-hand side are the standard ones used in holographic calculations in the supergravity approximation\footnote{In particular the extremal correlators among them vanish and the single-particle states are orthogonal to the multi-particle states.}. A signal that the microstate geometries are constructed in terms of $\gro{\cO^n}$ is that they induce a vev for dimension two operators~\cite{Giusto:2019qig,Rawash:2021pik} which would be absent when using the more standard definition because they involve extremal couplings. With a slight abuse of notation, from now on we will write between colons the (external) multi-trace operators relevant to the microstate geometries \eqref{eq:O:O:} and use the same notation for  \emph{correlators} involving this type of multi-trace operators. 
Thus we write the heavy-heavy-light-light correlation function on the plane\footnote{To go from correlators on the plane to the cylinder with our choice of light operator involves simply a Jacobian factor of $\abs{z}$.} derived in\footnote{In \cite{Bombini:2017sge} the result was given in terms of the continuous dimensionful parameters $a$, $a_0$, and $b$ which appear naturally in the supergravity description of the smooth geometry dual to the heavy state. 
Here it is more convenient to use a dimensionless parameter $B$, which is related to the supergravity parameters via
\begin{align*}
		\frac{a^2}{a_0^2}= 1 - \frac{B^2}{N}\,, \qquad \frac{b^2}{2a_0^2}=\frac{B^2}{N} \,,
\end{align*}
where we used the smoothness constraint $a^2 + b^2/2 = a_0^2$.} \cite{Bombini:2017sge} as
\begin{equation} \label{eq:fullcorr}
	\gro{C}\!(z, \zb) = \bigg(\!1-\frac{B^2}{N}\bigg) \sum_{k=1}^{\infty}\,\sum_{\ell\in \mathbb{Z}}\frac{1}{\sqrt{1- \frac{B^2}{N}\Big(1- \frac{\ell^2}{\left(|\ell|+2k\right)^2}\Big)}}\left(\frac{z}{\zb}\right)^{\frac{\ell}{2}} \left(z \zb\right)^{-\frac{|\ell|+2k}{2}\sqrt{1-\frac{B^2}{N}\big(1- \frac{\ell^2}{\left(|\ell|+2k\right)^2}\big)}}\,,
\end{equation}
which describes a correlator involving the heavy states
\begin{equation}\label{eq:O:H}
    \!:\mathcal{O}_{\!H\!,f}\!:\; = \sum_{p=0}^N \sqrt{\binom{N}{p}} \left(\frac{B}{\sqrt{N}}\right)^p \left(1-\frac{B^2}{N}\right)^{\frac{N-p}{2}}  \gro{\mathcal{O}^{\,p}_{\!f}}\;,
\end{equation}
in the regime where the parameter $B^2$ in chosen to scale with $N$ as $N\to\infty$. This scaling limit (the ``heavy scaling limit'') is necessary in order to treat the heavy operator as a smooth deformation of AdS$_3\times S^3$. However, following the discussion of Section~\ref{sec:bg} we are interested in the opposite regime, where the heavy operators are made light by keeping $B^2$ fixed and thus $B^2/N\to 0$ as $N\to\infty$. To describe this limit, it is convenient to rewrite the correlator \eqref{eq:fullcorr} as a series in $B^2/N$, as was done in \eqref{eq:HHLLseries}. We note that when $N$ is large and with $n$ finite, the binomial appearing in the last line of \eqref{eq:HHLLseries} can be approximated as 
\begin{align}
	\binom{N}{n} \xrightarrow[N\rightarrow\infty]{n \, \ll\, N} \frac{N^n}{n!}\,,
\end{align}
so that the expansion of the correlator can also be written as
\begin{align}
	\label{eq:serexpC}
	\gro{C}\!(z, \zb)\approx \sum_{n=0}^{\infty}\left(\frac{B^2}{N}\right)^{\!n} \, \frac{N^n }{n!}\,\gro{C_{n}}(z, \zb)\,.
\end{align}
This expression, together with the fact that connected correlators $\gro{C_{n}}$ scale as $N^{-n}$, makes it clear that the coefficients of each power of $B^2/N$ are finite in the large $N$ limit.

The main result of our analysis is that at each order in $B^2/N$ the double Fourier series in \eqref{eq:fullcorr} can be evaluated in a closed form using the procedure outlined in Appendix~\ref{app:corder} and one can rewrite the coefficients $\gro{C_{n}}(z, \zb)$ in terms of elementary functions and polylogarithms. 
In Appendix~\ref{app:corder} we present terms up to $n=4$, however, in the main text we limit ourselves to the first new result which occurs for $n=2$.
The for the first three values of $n$, the correlators are given by
\begin{subequations}
	\label{eq:corrs}
	\begin{align}
		{C_{0}} &= \frac{1}{|1-z|^2}\,, \label{eq:c0conn}\\
		{C_{1}} &= - \frac{\abs{z}^2}{N}\Biggr[\frac{4 i (z + \zb)}{(z- \zb)^3} P_2(z, \zb) + \frac{4}{(z - \zb)^2}\log|1-z|+ \frac{2(z + \zb - 2 z \zb)  }{(z- \zb)^2 |1-z|^2}\log|z|\Biggr]\nonumber\\*
		& \qquad \qquad - \frac{1}{N}\frac{1}{|1-z|^2}\,, \label{eq:c1conn}\\
		\gro{C_{2}} &= \frac{4\abs{z}^2}{N^2}\Biggr[\frac{6 i (z+ \zb) (z^2 + 10 z \zb +\zb^2) }{(z-\zb)^5} P_4(z ,\zb)  - \frac{12 (z^2 + 4 z \zb + \zb^2)}{(z-\zb)^4} P_3(z, \zb)\nonumber \\* 
		& \qquad \qquad+ \frac{8 i (z+ \zb)}{(z-\zb)^3}P_2(z, \zb) + \frac{2}{(z-\zb)^2}\, \log |1-z|+  \frac{z + \zb -2 z \zb}{(z- \zb)^2 |1-z|^2}\log|z|  \nonumber\\
		& \qquad\qquad+ \frac{z+ \zb}{|1-z|^2 (z-\zb)^2}\big(\log|z|\,\big)^2\Biggr]\,,\label{eq:c2conn}
	\end{align}
\end{subequations}
where the functions $P_n$ are the Bloch-Wigner dilogarithm and its generalisations (see for example \cite{zagier1990bloch}) discussed in section~\ref{sec:GBW}, see in particular~\eqref{eq:BWdef1} and~\eqref{eq:Pmdef}.

The result for $C_{0}$ is simply the identity contribution to the 4-point correlator and $C_{1}$ reproduces the results of \cite{Galliani:2017jlg, Bombini:2017sge}.
The first new expression is $\gro{C_{2}}$ which contains, in addition to the standard functions already appearing in known holographic correlators, the so-called Bloch-Wigner-Ramakrishnan (BWR) polylogarithm functions $P_m(z,\zb)$ \cite{zagier1991polylogarithms, zagier:2006dilog}, which for $m>2$ are higher order generalisations of the Bloch-Wigner dilogarithm. The same is true for the correlator $C_2$ involving the more standard double-particle operators defined as the OPE limit of mutually BPS single-particle operators. From the general structure shown in~\eqref{eq:O:O:}, one can see that the difference between $C_2$ and $\gro{C_2}$ can be written in terms of connected $k$-point correlators with $k\leq 4$ and so these contributions do not affect the terms proportional to $P_4$ and $P_3$ in~\eqref{eq:c2conn}. A similar structure can be found in the correlators $\gro{C_{n}}$ for larger values of $n$, with the highest order BWR polylogarithm function appearing being $P_{2n}$, as we show for some additional cases in Appendix~\ref{app:corder}. This contribution is not sensitive to the differences between $\gro{C_{n}}$ and $C_n$ showing that the presence of the higher order BWR polylogarithm functions is a generic feature of the OPE limit of supergravity holographic correlators with many single-particle external states.

\subsection{Generalised Bloch-Wigner Functions}
\label{sec:GBW}

The expression for $\gro{C_{2}}$ in \eqref{eq:c2conn} (and similarly for the higher order counterparts, see \eqref{eq:Bcor-short}) involves Bloch-Wigner-Ramakrishnan polylogarithms -- analogues of the standard Bloch-Wigner functions containing higher-order polylogarithms -- and since these generalised functions are not widely used in the literature, we use this subsection to briefly review their definition and some of their properties.
In doing this we closely follow the works of Lewin \cite{Lewin:100000}, and Zagier~\cite{zagier:2006dilog} who first presented the generalisation of the Bloch-Wigner function in \cite{zagier1990bloch, zagier1991polylogarithms} following the work of Ramakrishnan \cite{Ramakrishnan:1986}.

The starting point of our analysis is the polylogarithm function, which admits the series representation 
\begin{align}
	\label{eq:plogdef}
	\plog{n}{z} = \sum_{k=1}^{\infty} \frac{z^k}{k^n}\,, \qquad \abs{z} < 1\,,
\end{align}
where we take $z$ to be a complex number and $n$ an integer.
For non-positive integer $n$ this defines a rational function, while for $n=1$ it gives simply
\begin{align}
	\label{eq:pl1}
\plog{1}{z} = - \log{(1-z)}\,.
\end{align} 
Polylogarithm functions associated with higher integer values do not admit a representation in terms of elementary functions but can be obtained by recursion relations that follow from their definition	
\begin{align}
	\label{eq:polylogrec}
	z \,\partial_z \plog{n}{z} = \plog{n-1}{z}\,,\qquad 
	\plog{n+1}{z} = \int_{0}^{z} \frac{\plog{n}{w}}{w}\, dw\,,
\end{align}
which can also be used to analytically continue the polylogarithm functions to generic $z \in \mathbb{C}\setminus [1, \infty)$ by repeated integration starting from \eqref{eq:pl1}. 
Therefore, the polylogarithm functions inherit some of their properties from the logarithm such as a branch cut along the real axis starting at $z = 1$. 

The dilogarithm function $\plog{2}{z}$ has played an important role in the CFT literature, as it appears in the expressions of holographic 4-point correlation functions of single-trace operators. 
In fact, is it not the ``bare'' dilogarithm function that appears in these correlators but a particular combination called the Bloch-Wigner dilogarithm function, defined as%
\footnote{It is customary to denote the Bloch-Wigner dilogarithm function by $D(z, \zb)$. Here we denote it with $P_2(z, \zb)$ in order to facilitate an easier generalisation to higher order functions.}
\begin{align} \label{eq:BWdef1}
	P_2(z, \zb)& \equiv  \mathrm{Im}\big[\plog{2}{z}\!\big] + \log\abs{z}\, \arg\big[1-z\big]\nonumber\\*
	& =  \frac{1}{2i}\left[ \plog{2}{z} - \plog{2}{\zb} +  \log\abs{z}\, \log\left( \frac{1-z}{1-\zb}\right)\right]\,, 
\end{align}
which has some advantageous properties compared with the ordinary dilogarithm function.
For example, the dilogarithm function satisfies various identities relating the values of the function at different points in the complex plane, such as
\begin{subequations}
	\label{eq:pl2rel}
	\begin{align}
		\plog{2}{\frac{1}{z}} & = - \plog{2}{z} - \frac{\pi^2}{6} - \frac12 \log^2 \left(-z\right)\,,\\
		 \plog{2}{1-z} &= - \plog{2}{z} + \frac{\pi^2}{6}- \log(z)\log(1-z)\,.
	\end{align}
\end{subequations}
In fact, one can show that the values of $\plog{2}{z}$,  $\plog{2}{\frac{1}{1-z}}$, $\plog{2}{\frac{z-1}{z}}$, $-\plog{2}{\frac{1}{z}}$, $-\plog{2}{\frac{z}{1-z}}$, and $-\plog{2}{1-z}$ are all equal up the addition of elementary functions such as logarithms, as in \eqref{eq:pl2rel}. 
In contrast to this, the function defined in \eqref{eq:BWdef1} satisfies similar identities but importantly without any additional elementary function terms
\begin{align}
	\label{eq:Dsym}
	P_2(z, \zb) &= 	-P_2\left(\frac1{z}, \frac1{\zb}\right)  = - P_2(1-z, 1-\zb)= P_2\left( \frac{z-1}{z}, \frac{\zb-1}{\zb}\right)\nonumber \\
	&= P_2\left( \frac{1}{1-z}, \frac{1}{1-\zb}\right)= - P_2\left( \frac{z}{z-1}, \frac{\zb}{\zb-1}\right)\,.
\end{align}
Considering the many symmetries of 4-point correlation functions involving single-trace operators, such symmetry properties are not unexpected. 
Furthermore, if $\zb= z^*$ the Bloch-Wigner function is real-analytic on the whole complex plane bar the points $z =0$ and $z =1$ where it is continuous but not differentiable. 
Thus one finds no branch cut, unlike in the case of the dilogarithm.
In our analysis we often take the analytically continued function where $z$ and $\zb$ are independent, in which case a more complicated analytic structure emerges \cite{Maldacena:2015iua}. 
Contrary to the dilogarithm, a generic polylogarithm function satisfies fewer simple symmetry identities. 
In fact, the inversion identity relating $\plog{n}{z}$ and $(-1)^{n-1}\plog{n}{\frac1{z}}$ (up to several terms that include different powers of $\log{z}$) is in general the only functional identity that involves polylogarithm functions evaluated at two generic points in the complex plane, while other identities typically relate the values of polylogarithm functions evaluated at multiple points.
Nonetheless, it is possible to define a higher-order generalisation of \eqref{eq:BWdef1} as \cite{zagier1991polylogarithms}
\begin{align}
	\label{eq:Pmdef}
	P_n(z, \zb) &= \mathfrak{R}_n \left(\, \sum_{k = 0}^{n-1} \frac{ 2^k\,B_k}{k!}\, \big( \log|z|\,\big)^k\, \plog{n-k}{z}\right)\,,
\end{align}
where $\mathfrak{R}_n$ denotes the real or imaginary part of the expression when $n$ is odd or even respectively and the coefficients $B_j$ denote the Bernoulli numbers%
\footnote{The first few non-zero Bernoulli numbers are $B_0 = 1$, $B_1 = -\frac12$, $B_2 = \frac16$, $B_4 = - \frac1{30}$, $B_6 = \frac{1}{42}$, $B_8 =  -\frac{1}{30}$. Apart from $B_1$ all Bernoulli numbers with an odd index vanish.}.
Once again, as a function of a single complex variable (when $\zb= z^*$), $P_n(z, \zb)$ defines a real-analytic function on the complex plane, except at $z=0$ and $z=1$ where is it only continuous. 
In addition, the functions obey the inversion relation
\begin{align}
	P_n(z, \zb) = (-1)^{n-1} P_n\left( \frac{1}{z}, \frac{1}{\zb}\right) \,,
\end{align}
however, the higher order generalisations do not obey other simple symmetry identities found in the $n=2$ case \eqref{eq:Dsym}. 
For completeness, we give the explicit expressions of the first few generalised Bloch-Wigner-Ramakrishnan functions, which are explicitly used in the main text
\begin{subequations}
	\label{eq:Pdef}
	\begin{align}
		P_2(z,\zb) & =  \frac{1}{2i}\bigg[ \plog{2}{z} - \plog{2}{\zb} + \log|z|\, \log\left( \frac{1-z}{1-\zb}\right) \bigg]\,,\\
		P_3(z,\zb) &= \frac12 \bigg[ \plog{3}{z} + \plog{3}{\zb} - \log|z|\, \Big(\plog{2}{z} + \plog{2}{\zb}\!\Big) - \frac{2}{3} \big( \log|z|\,\big)^2\, \log|1-z| \bigg]\,,\\
		P_4(z,\zb) &= \frac{1}{2i}\bigg[ \plog{4}{z} - \plog{4}{\zb} - \log|z|\,\Big( \plog{3}{z} - \plog{3}{\zb}\!\Big) \nonumber\\
		&\hspace{6.6cm}+ \frac{1}{3}\big(\log|z|\,\big)^2 \Big( \plog{2}{z} - \plog{2}{\zb}\!\Big)\bigg]\,.
	\end{align}
\end{subequations}
Further examples can be found in Appendix~\ref{app:corder}.

\section{OPE limits}
\label{sec:lchecks}

We proposed to identify the $B^{2n}$ term in the expansion of the HHLL correlator \eqref{eq:fullcorr} computed in \cite{Bombini:2017sge} with the ``connected" correlator $\gro{C_{n}}$ containing two $n$-trace operators of the type $\gro{\mathcal{O}^{\,n}_{\!f}}$ and two single-trace operators. In this section we provide evidence supporting this identification, concentrating on the first non-trivial example -- that of $n=2$. As a first check of this result we focus on protected terms in the OPE expansion of the correlator~\eqref{eq:c2conn} that appear in the $z\to 1$ and $z\to 0$ channels. We show that these terms match the results obtained from CFT calculations at the free orbifold point, as required by non-renormalisation theorems~\cite{Baggio:2012rr} and by the affine symmetry.

We then look at terms originating from the exchange of non-protected double-trace operators appearing in the $z\to 1$ OPE. By following~\cite{Freedman:1998bj}, we find the anomalous dimensions and 3-point couplings of the non-protected double-trace operators from $n=1$ correlators. With respect to the case of ${\cal N}=4$ SYM~\cite{Aprile:2017xsp}, there is an extra complication~\cite{Aprile:2021mvq} related to the flavour symmetry of the CPOs (in passing we point out that the theory we are studying provides a top-down example of the pattern highlighted in~\cite{Alday:2017gde}, where it was noted, in an effective field theory approach, that anomalous dimensions of double-trace operators with low spin can be positive). Due to the flavour mixing one would need results for other correlators, besides~\eqref{eq:fullcorr}, in order to extend this analysis to $n>1$ and extract precise information about the multi-trace operators exchanged. Here we focus on general checks relating to the presence or absence of $\log$ and $\log^2$ terms in the $z\to 1$ OPE. The $z\to\infty$ OPE involves triple-trace operators even at the first non-trivial order of the $n=2$ correlator and again one would need other correlators to disentangle the flavour dependence and extract CFT data at strong coupling. We leave a detailed study of the multi-trace anomalous dimensions and 3-point couplings in the AdS$_3$/CFT$_2$ case to a future analysis.

\subsection{The protected sector}

In the limit where $\bar z\to 1$ with $z$ kept fixed, the leading term of the correlator is determined by the exchange of operators with right conformal dimension $\bar h=0$ and generic left conformal dimension $h$. These are the chiral currents of the theory, whose modes are the Virasoro $L_{-n}$ and the R-symmetry\footnote{Only R-symmetry neutral operators contribute in the $z\to 1$ channel, and thus we can restrict to the $U(1)$ subgroup of the $SU(2)$ R-symmetry.} $J^3_{-n}$ generators. Since their 3-point couplings are determined by symmetries of the theory, they can be computed exactly at any point in the CFT moduli space and compared with the gravity result.

The $z, \bar z\to 0$ limit of the correlator $\gro{\mathcal{C}_n}$ is controlled by the non-singular OPE of the operators $\mathcal{O}_{\!g}$ and $\gro{\mathcal{O}_{\!f}^{\,n}}$ -- chiral primary operators preserving the same supercharges. The lowest dimension operator exchanged in this channel is thus protected and we reproduce the vanishing of the term of order $z^0\zb^0$ in $\gro{C_{2}}$ from a computation at the free orbifold point.

\subsubsection{$\bar z\to 1$ light-cone OPE}

This check of the protected current contributions to the tree-level connected correlators \eqref{eq:corrs} can be simplified slightly by replacing the two single-trace operators $\mathcal{O}_{\!g}\,$, $\bar{\mathcal{O}}_{\!g}$ with their superdescendant $\mathcal{O}^B_{\!g}$ -- obtained by acting on $\mathcal{O}_{\!g}$ with a right-moving and a left-moving supercharge. The corresponding correlator, which we will denote by $\gro{\mathcal{C}^B_n}$, is related to $\gro{\mathcal{C}_n}$ by the supersymmetry Ward identity \cite{Bombini:2017sge}
\begin{equation}\label{eq:WI}
	\gro{\mathcal{C}^B_n}\!(z,\bar z) = \partial\bar\partial\, \gro{\mathcal{C}_n}\!(z,\bar z)\,.
\end{equation}
Starting from the correlator $\gro{C_{2}}$ in \eqref{eq:c2conn}, applying the Ward identity \eqref{eq:WI} and taking the leading order term as $\bar z\to 1$ one obtains 
\begin{equation}\label{eq:gravityresult}
	\gro{C^B_{2}}\!(z,\bar z)\stackrel{\bar z \to 1}{\longrightarrow} \frac{\gro{G^B_{2}}\!(z)}{|1-z|^4}\ \text{ with }\ N^2\! \gro{G^B_{2}}\!(z) =2+4 \frac{1+z}{1-z}\log z + \frac{1+4z+z^2}{(1-z)^2}\big(\log z\big)^2\,.
\end{equation}
The goal of this subsection is to reproduce $\gro{G^B_{2}}$ from a CFT computation. Since $\mathcal{O}^B_{\!g}$ has vanishing R-charge, the states obtained by acting on the vacuum with the $J^3_{-n}$ modes of the current do not contribute to $\gro{\mathcal{C}^B_n}$. In order to decouple the Virasoro from the R-current algebras it is convenient to subtract from the $L_{-n}$ modes the Sugawara contribution: the algebra satisfied by these ``reduced" Virasoro generators, which for notational simplicity we will still denote by $L_{-n}$, is identical to the Virasoro algebra but the conformal dimension $h$ of an operator of R-charge $j$ should be replaced by a ``reduced" conformal dimension $h^{[0]}\equiv h-\frac{j^2}{N}$.

States of the form $L_{-n_1}L_{-n_2}\dots L_{-n_p}|0\rangle$, containing $p$ modes, have a norm proportional to $c^p\sim N^p$ and thus contribute to a correlator at order $N^{-p}$. Since we focus on $\gro{C^B_{2}}$ whose tree-level contribution is of order $N^{-2}$, we can simply consider such states up to $p=2$. Resumming the states with $p=1$ gives the well-known global conformal block of the stress-tensor: for a correlator $\mathcal{C}$ with two operators of dimension $h_g$ and two operators of dimension $h_f$, the contribution of these states to $\mathcal{G}=\abs{1-z}^{4h_g}\mathcal{C}$ is 
\begin{equation}
	p=1:\quad \frac{1}{N} \,h_f h_g \,\mathcal{V}_1(1-z)\quad \mathrm{with}\quad \mathcal{V}_1(z)=\frac{1}{3} z^2 \,{}_2F_1(2,2,4;z)\,.
\end{equation}
Likewise, the contribution due to the exchange of states with $p=2$ to this same correlator is a sum of terms of the form
\begin{equation}
	p=2: \quad \frac{1}{N^2} \left(h_f^2 h_g^2 \,\mathcal{V}^{(2,2)}_2(1-z)+ (h_f^2 h_g+ h_f h_g^2)\, \mathcal{V}^{(2,1)}_2(1-z) + h_f h_g \, \mathcal{V}^{(1,1)}_2(1-z)\right) \,.
\end{equation}
The functions $\mathcal{V}^{(2,2)}_2$ and $\mathcal{V}^{(2,1)}_2$ can be readily computed and are given in Appendix D of \cite{Fitzpatrick:2015qma} as
\begin{equation}\label{eq:V22V21}
	\begin{aligned}
		\mathcal{V}^{(2,2)}_2(z) &=\frac{1}{18} \big(z^2 {}_2F_1(2,2,4;z)\big)^2\,, \\
		\mathcal{V}^{(2,1)}_2(z) &=-\frac{1}{18} \left(\big(z^2 {}_2F_1(2,2,4;z)\big)^2 +\frac{6}{5}\,z^3 \log(1-z)\,{}_2F_1(3,3,6;z)\right)\,,
	\end{aligned}
\end{equation}
while, as will be clear in a moment, we will not need $\mathcal{V}^{(1,1)}_2$. For the correlator $\gro{\mathcal{C}^B_n}$ we should take $h_g=1$, the dimension of the superdescendant $\mathcal{O}^B_{\!g}$, and $h_{f,n}=\frac{n}{2}- \frac{n^2}{4\,N}$, the reduced dimension of the $n$-trace operator $\gro{\mathcal{O}^{\,n}_{\!f}}$, where we made the dependence on $n$ explicit. Then combining the correlators $\gro{\mathcal{C}^B_1}$ and $\gro{\mathcal{C}^B_2}$ according to the definition of the connected combination $\gro{C^B_{2}}$ (of the same form as~\eqref{eq:n=2ex}) we deduce that in the light-cone limit and up to terms of order $1/N^3$
\begin{align}\label{eq:N2lope}
    \gro{G^B_{2}}\!(z) &= \frac{1}{N} \big( h_{f,2} - 2 h_{f,1} \big)\mathcal{V}_1(1-z) +\frac{1}{N^2} \bigg[ \big(h^2_{f,2} - 2 h^2_{f,1}\big)\mathcal{V}^{(2,2)}_2(1-z) \nonumber\\ 
    &\ \ \ +\Big( \big(h^2_{f,2}+h_{f,2}\big) - 2 \big(h^2_{f,1}+h_{f,1}\big)\Big) \mathcal{V}^{(2,1)}_2(1-z) +\big(h_{f,2} - 2 h_{f,1}\big)\mathcal{V}^{(1,1)}_2(1-z) \bigg] \nonumber\\ 
    &=\frac{1}{N^2}\left[2+4 \frac{1+z}{1-z}\log z + \frac{1+4z+z^2}{(1-z)^2}\big(\log z\big)^2 \right]+ O\!\left(\frac{1}{N^3}\right) \,,
\end{align}
where in the first step we used $h_g=1$ and in the second we used~\eqref{eq:V22V21} along with the large $N$ expansion of the $h_{f,n}$. The final result agrees with the expression \eqref{eq:gravityresult} obtained from the gravity computation. We note that in taking the connected combination both the $1/N$ term and the term proportional to $\mathcal{V}^{(1,1)}_2$ cancel. This latter term is interpreted as a quantum contribution related to bulk diagrams of the type shown in c) of Fig.~\ref{Fig:WDB4} which vanish in the limit that a pair of operators is made heavy (i.e. $h_f\sim N$) and $N$ is taken to infinity. One can show that analogous cancellations happen with $\gro{C^B_{n}}$ for generic $n$.

\subsubsection{Euclidean $z,\bar{z}\to 0$ OPE}

The exchanged operator with lowest dimension in this channel is the supersymmetric multi-trace $:\!\mathcal{O}_{\!f}^{\,n} \mathcal{O}_{\!g}\!:$ whose protected 3-point couplings can be computed in the orbifold sigma-model. One can thus verify the gravity prediction, which gives for the connected correlator $\gro{C_{2}}$ a vanishing coefficient for the lowest order term $z^0 \bar z^0$:
\begin{equation} \label{eq:C2connzzb0}
	\gro{C_{2}}\, \xrightarrow{z,\bar z\to 0} O(z \bar z)\,. 
\end{equation}
The explicit expressions of the relevant single- and multi-trace operators in the $\mathcal{M}^N/S_N$ orbifold theory are 
\begin{align}\label{eq:FreeOpDefs}
	&\qquad\qquad\qquad\mathcal{O}_{\!f}=\frac{1}{\sqrt{N}}\sum_r \mathcal{O}_{\!f (r)}\,,\quad :\!\mathcal{O}^{\,2}_{\!f}\!:\,=\frac{1}{\sqrt{\binom{N}{2}}}\sum_{r<s} \mathcal{O}_{\!f (r)} \mathcal{O}_{\!f(s)}\,,\\
	& :\!\mathcal{O}_{\!f} \mathcal{O}_{\!g}\!:\,=\frac{1}{\sqrt{N(N-1)}}\sum_{r\not=s}\mathcal{O}_{\!f (r)} \mathcal{O}_{\!g (s)}\,,\quad :\!\mathcal{O}^{\,2}_{\!f} \mathcal{O}_{\!g}\!:\,=\frac{1}{\sqrt{\binom{N}{2}(N-2)}}\sum_{r<s \atop t\not=r,s} \mathcal{O}_{\!f (r)} \mathcal{O}_{\!f(s)} \mathcal{O}_{\!g (t)}\,, \nonumber
\end{align}
where the subscripts $(r), (s), \ldots$ denote the $N$ copies of $\mathcal{M}$. The operators $\mathcal{O}_{\!f(r)}$ on different copies are orthogonal
\begin{equation}
	\langle \mathcal{O}_{\!f (r)} \mathcal{O}_{\!g (s)}\rangle = \delta_{f,g} \delta_{r,s}\ ,
\end{equation}
and the $N$-dependent prefactors in \eqref{eq:FreeOpDefs} have been chosen to normalise two-point functions to 1. The relevant 3-point functions can then be immediately computed as
\begin{equation}\label{eq:couplinz0}
	\langle \bar{\mathcal{O}}_{\!f}  \bar{\mathcal{O}}_{\!g} :\!\mathcal{O}_{\!f} \mathcal{O}_{\!g}\!:\rangle=\left(1-\frac{1}{N}\right)^{\frac{1}{2}}\,,\quad \langle :\!\bar{\mathcal{O}}_{\!f}^{\,2}\!\!:  \bar{\mathcal{O}}_{\!g} :\!\mathcal{O}^{\,2}_{\!f} \mathcal{O}_{\!g}\!:\rangle=\left(1-\frac{2}{N}\right)^{\frac{1}{2}}\,,
\end{equation}
and the coefficient of the $z^0 \bar z^0$ term in $\gro{C_{2}}$ is
\begin{equation}
	\left.\gro{C_{2}}\right|_{z^0\!,\bar z^0}=\big|\langle :\!\!\bar{\mathcal{O}}_{\!f}^{\,2}\!\!: \bar{\mathcal{O}}_{\!g} :\!\!\mathcal{O}^{\,2}_{\!f} \mathcal{O}_{\!g}\!\!:\rangle\big|^2- 2\,\big|\langle \bar{\mathcal{O}}_{\!f} \bar{\mathcal{O}}_{\!g} :\!\!\mathcal{O}_{\!f} \mathcal{O}_{\!g}\!\!:\rangle\big|^{2} +1=\left(1-\frac{2}{N}\right)-2\left(1-\frac{1}{N}\right)+1=0\,;
\end{equation}
the cancellation between the three terms, which come respectively from $\gro{\mathcal{C}_2}$, $-2\mathcal{C}_1$ and $\mathcal{C}_0$, is thus in agreement with the gravity prediction \eqref{eq:C2connzzb0}.

\subsection{The non-protected sector of the $z,\bar{z}\to 1$ OPE}

In the $z,\bar z\to 1$ OPE channel the dynamical contribution is due to the exchange of non-supersymmetric double-trace operators made of equal-flavour and opposite R-symmetry constituents, of the type $:\!\mathcal{O}_{\!f} \bar{\mathcal{O}}_{\!f}\!:$. At leading order in $N$, these operators are degenerate with the more general class of double-traces $:\!\mathcal{O}^{(\alpha,\dot\alpha)}_{\!f} \bar{\mathcal{O}}^{(\beta,\dot\beta)}_{\!g}\!:$ and in order to see how this degeneracy is lifted by $1/N$ corrections one needs to introduce a proper basis in the R-symmetry and flavour space. As a first step, we separate the results into the irreducible R-symmetry representations of the operators exchanged, which we characterise by their R-charge $(j,\bar{j})$. In the $n=1$ case enough information is available in the literature to perform a similar decomposition also in flavour space for all representations\footnote{At present this is true for correlators where the external operators have dimension $h=\bar{h}=1/2$ and, as discussed in~\cite{Aprile:2021mvq}, for generic dimensions in the case of non-singlet flavour representations.} and hence in the first subsection we derive the ``unmixed'' CFT data for the relevant anomalous dimensions. In the second subsection we focus on the $n=2$ correlator~\eqref{eq:c2conn} but limit ourselves only to averaged CFT data, however, we point out that this is sufficient to obtain a further consistency check of our results.

Though the results in~\eqref{eq:c0conn} and~\eqref{eq:c1conn} refer to correlators containing operators of two different flavours and with the particular R-symmetry choice obtained by replacing $\mathcal{O}_{\!f}$ in~\eqref{eq:multiTcorr} with the highest R-charge operator $\mathcal{O}_{\!f} \equiv \mathcal{O}^{++}_{\!f}$ defined in~\eqref{eq:OLexp}, the analysis of the following subsections would require a more general class of correlators%
\footnote{To avoid clutter, we suppress the $z$ and $\zb$ dependence of correlators such as \eqref{eq:Cabfi} throughout this section. It should be understood that operators are always inserted as in \eqref{eq:defCn}.}
\begin{equation} \label{eq:Cabfi}
    \gro{\mathcal{C}^{\alpha\dot{\alpha},\,\beta\dot{\beta}}_{n, f_1 f_2 f_3 f_4}}\; = \langle\, \gro{\!(\mathcal{O}^{--}_{\!f_1})^n\!}\, \gro{\!(\mathcal{O}^{++}_{\!f_2})^n\!}\, \mathcal{O}^{\alpha\dot\alpha}_{\!f_3}\, \mathcal{O}^{\beta\dot\beta}_{\!f_4}\, \rangle \ .
\end{equation}
These are known for generic flavour and R-symmetry indices only for $n=1$, whose expressions at order $1/N$ we summarise in Appendix~\ref{app:n1summ}. 

To carry out the program outlined above and separate the different irreducible representations in the flavour and R-symmetry space, we need to introduce the appropriate projectors. For the flavour part, the exchanged operator in the $z\to 1$ Euclidean OPE sits in the product of two fundamental $SO(N_f)$ representations and can be decomposed in the singlet, symmetric-traceless and anti-symmetric irreps, whose contributions to \eqref{eq:Cabfi} can be selected by using the projection operators
\begin{subequations}
\label{eq:projectors}
    \begin{align} \label{eq:projectors1}
        \mathcal{P}^{\mathrm{sing}}_{\!f_3f_4,\,g_3g_4} &= \frac1{N_f}\delta_{f_3f_4}\delta_{g_3g_4} \ , \\ \label{eq:projectors2}
        \mathcal{P}^{\mathrm{sym}}_{\!f_3f_4,\,g_3g_4} &= \frac1{2}\Big( \delta_{f_3g_3}\delta_{f_4g_4}+\delta_{f_3g_4}\delta_{f_4g_3} - \frac2{N_f}\delta_{f_3f_4}\delta_{g_3g_4}\Big) \ , \\\label{eq:projectors3}
        \mathcal{P}^{\mathrm{asym}}_{\!f_3f_4,\,g_3g_4} &= \frac1{2}\Big( \delta_{f_3g_3}\delta_{f_4g_4}-\delta_{f_3g_4}\delta_{f_4g_3}\Big) \ .
    \end{align}
\end{subequations}
Similarly, for the R-symmetry we can separate the singlet ($j=0$) and the triplet ($j=1$) in the product of the two $SU(2)_L$ doublets $(\alpha,\beta)$ using the projectors
\begin{align} \label{eq:Rprojectors}
   \mathcal{R}^{\alpha\beta,\gamma\delta}_{(j)} =\frac{1}{2} \,\sigma_{(j)}^{\alpha\beta}\, \sigma_{(j)}^{\gamma\delta}\ \text{ where }\  \sigma_{(0)}^{\alpha\beta}=\begin{pmatrix} 0&-i\\i&0\end{pmatrix}\ \text{ and }\  \sigma_{(1)}^{\alpha\beta}=\begin{pmatrix}0&1\\1&0\end{pmatrix}\,.
\end{align}
Identical projectors $\mathcal{R}_{\alpha\beta,\gamma\delta}^{(\jb)}$ with $\jb=0,1$ act on the $SU(2)_R$ indices $(\dot\alpha,\dot\beta)$. The correlator \eqref{eq:Cabfi} can then be decomposed into its irreducible R-symmetry and flavour components $\gro{\mathcal{C}^{\mathrm{flav}}_{n \,(j,\jb)}}$ as
\begin{equation} \label{eq:irreps}
    \gro{\mathcal{C}^{\alpha\dot{\alpha},\,\beta\dot{\beta}}_{n, f_1 f_2 f_3 f_4}}\; =\sum_{j,\jb=0,1}\sum_{\mathrm{flav}} \sigma_{(j)}^{\alpha\beta} \,\sigma_{(\jb)}^{\dot\alpha\dot\beta} \,\mathcal{P}^{\mathrm{flav}}_{f_1 f_2 f_3 f_4} \gro{\mathcal{C}^{\mathrm{flav}}_{n \,(j,\jb)}}\ ,
\end{equation}
where $\mathrm{flav}=\mathrm{sing}, \mathrm{sym}, \mathrm{asym}$. Given the linear relation \eqref{eq:defconn} between the correlators $\gro{\mathcal{C}_n}$ and their connected combinations $\gro{C_n}$, an identical decomposition defines the irreducible components $\gro{C^{\mathrm{flav}}_{n \,(j,\jb)}}$ of the connected correlators.

\subsubsection{The $n=1$ case and the double-trace CFT data}\label{sssec:n1}

In this subsection we focus on the $n=1$ correlators, which we give for arbitrary values of the flavour and the R-symmetry indices in Appendix~\ref{app:n1summ}, extracting the OPE data for the lowest twist double-trace operators exchanged in the $z,\bar{z} \to 1$ channel. As standard in the Euclidean OPE, we can expand each function $\mathcal{C}^{\mathrm{flav}}_{1\,(j,\jb)}$ defined in \eqref{eq:irreps} as $z,\bar{z} \to 1$ and rewrite the result in terms of {\it global} conformal blocks. The coefficients of this block decomposition admit a large $N$ expansion of the form
\begin{equation}\label{eq:n1datacoeff}
    \abs{c^{\mathrm{flav},k}_{(0)(j,\jb)}}^2 + \frac1{N} \abs{c^{\mathrm{flav},k}_{(1)(j,\jb)}}^2 + \frac1{N^2} \abs{c^{\mathrm{flav},k}_{(2)(j,\jb)}}^2 + O\big(N^{-3}\big)\,.
\end{equation}
The leading order coefficients $\big|c^{\mathrm{flav},k}_{(0)(j,\jb)}\big|^2$ are explained by the exchange of non-BPS double-trace operators of the form $(\mathcal{O}\bar{\mathcal{O}})_{k}\;\sim\; \mathcal{O}_{\!f_3}\pd^k\bar{\mathcal{O}}_{\!f_4}$ with conformal dimensions
\begin{equation}
\begin{aligned}\label{eq:n1datadim}
	h_{(j,\jb)}^{\mathrm{flav},k}&= 1+k + \gamma_{(j,\jb)}^{\mathrm{flav},k} \approx 1+k+\frac1{N} \gamma_{(1)(j,\jb)}^{\mathrm{flav},k} + \frac1{N^2} \gamma_{(2)(j,\jb)}^{\mathrm{flav},k} \ , \\
	\hb_{(j,\jb)}^{\mathrm{flav},k} &= 1 + \gamma_{(j,\jb)}^{\mathrm{flav},k} \approx 1+\frac1{N} \gamma_{(1)(j,\jb)}^{\mathrm{flav},k} + \frac1{N^2} \gamma_{(2)(j,\jb)}^{\mathrm{flav},k} \ .
\end{aligned}
\end{equation}
Then one has the identification
\begin{align}\label{eq:n1data}
	\langle\bar{\mathcal{O}}_{\!f_1}\mathcal{O}_{\!f_2}(\mathcal{O}\bar{\mathcal{O}})_{k}\rangle\langle(\mathcal{O}\bar{\mathcal{O}})_{k}\mathcal{O}^{(\alpha,\dot\alpha)}_{\!f_3}\mathcal{O}^{(\beta,\dot\beta)}_{\!f_4}\rangle \mathcal{P}^{\mathrm{flav}}_{\!f_1f_2,\,f_3 f_4}= \abs{c^{\mathrm{flav},k}_{(0)(j,\jb)}}^2\sigma^{\alpha\beta}_{(j)} \sigma^{\dot\alpha\dot\beta}_{(\bar j)} +O(N^{-1})\,.
\end{align}
The subleading terms $\big|c^{\mathrm{flav},k}_{(1)(j,\jb)}\big|^2$, $\big|c^{\mathrm{flav},k}_{(2)(j,\jb)}\big|^2$ receive contributions both from the $1/N$ and $1/N^2$ corrections to the 3-point couplings in \eqref{eq:n1data} and from the couplings with triple-trace and quadruple-trace operators, which start at order $1/N$ and $1/N^2$ respectively. We will not try to disentangle these different contributions, but simply extract the total $1/N$ coefficients $\big|c^{\mathrm{flav},k}_{(1)(j,\jb)}\big|^2$, which are also the data computable from the inversion formula (as done in Appendix~\ref{app:inversion}). 

However, the $n=1$ correlators summarised in Appendix~\ref{app:n1summ} are sufficient to derive the anomalous dimensions of the true conformal primaries at order $O(1/N)$ for each value of $k$. Let us start from the contribution related to double-trace operators in the R-symmetry singlet irrep $(j,\jb)=(0,0)$ and flavour singlet: we expand $C^{\mathrm{sing}}_{1\,(0,0)}$ up to order $(1-\bar{z})^0$ and keep the order of $(1-z)^t$ arbitrary and obtain
\begin{align} \label{eq:s1sisik}
	C^{\mathrm{sing}}_{1\,(0,0)} \approx &\sum_{t=0}^{\infty}\, (1-z)^t\bigg[ \frac{1}{2N}\bigg(\frac{(t^2+t+2)N_f}{(t+1)(t+2)(t+3)} -\big(1+\delta_{t,0}\big)\bigg)\log\abs{1-z}^2 + \frac{1+\delta_{t,0}}{4}\nonumber \\*
	&\quad\ + \frac{1}{N}\Big(A_t N_f+B_t\Big)\bigg] + \sum_{t=0}^{\infty}\frac{(-1)^t(1-z)^{t+2}}{N\abs{1-z}^2}\frac{(t+1)N_f}{2(t+2)(t+3)} \ ,
\end{align}
where the first values of $A_t$ and $B_t$ are
\begin{equation}\label{eq:AkBk}
    \begin{aligned}
        A_t &= \bigg\{\! -\frac{7}{36}\,, -\frac1{72}\,, -\frac1{225}\,,-\frac1{900}\,, \frac{61}{44100}\,, \frac{23}{7056}\,, \frac{293}{63504}\,, \dots\bigg\}\;, \\
        B_t &= \bigg\{\frac1{2}\,,-\frac1{4}\,,-\frac1{6}\,,-\frac1{6}\,,-\frac{7}{40}\,,-\frac{11}{60}\,,-\frac{4}{21}\,,\dots\bigg\} \ .
    \end{aligned}
\end{equation}
By focusing on the leading term in $N$, we can extract~\cite{Heemskerk:2009pn} a closed expression for the OPE coefficients with a double-trace operator of dimension $(h,\bar{h})=(k+1,1)+O(1/N)$
\begin{equation} \label{eq:c0ksingsing}
    \abs{c^{\mathrm{sing},k}_{(0)(0,0)}}^2 = \Big(1+(-1)^k\Big)\frac{(k!)^2}{4(2k)!} \ .
\end{equation}
As expected, only states with even spin are non-trivial. Likewise, by projecting along the conformal blocks, from the term proportional to $\log|1-z|^2$ we obtain the anomalous dimensions
\begin{equation} \label{eq:delksing}
    \abs{c^{\mathrm{sing},k}_{(0)(0,0)}}^2 \gamma_{(1)(0,0)}^{\mathrm{sing},k} = \frac{(k!)^2}{2(2k)!}\bigg[\frac{(-1)^k(k^2+k-2)^2}{\Gamma(k+4)\Gamma(3-k)}N_f - \Big(1+(-1)^k\Big)\bigg] \ .
\end{equation}
We note that for $k>2$ the first term in the square parenthesis vanishes (due to the $\Gamma(3-k)$ in the denominator) and the result for $\gamma_{(1)}$ agrees with that obtained from the Lorentzian inversion relation~\cite{Caron-Huot:2017vep}, as seen for instance in~\cite{Kraus:2018zrn} and in \eqref{eq:nkCFTDatag} with $\nb=0$. Then the pattern of these anomalous dimensions is
\begin{equation} \label{eq:AnomDimSing}
    \gamma_{(1)(0,0)}^{\mathrm{sing},k} = \bigg\{ \frac{N_f}{3}-2, 0,\frac{2N_f}{15}-2,0, -2,0,-2,0,\dots\bigg\} \ .
\end{equation}
In the case of AdS$_3 \times S^3\times T^4$, we have $N_f=5$ and all $\gamma_{(1)(0,0)}^{\mathrm{sing},k}$ are negative down to $k=0$: this is also the case in ${\cal N}=4$ SYM~\cite{Alday:2017vkk}. However, for AdS$_3 \times S^3\times K3$, one has $N_f=21$ and so the first two non-trivial values in \eqref{eq:AnomDimSing}, {\rm i.e.} $k=0,2$, are positive. This realises the possibility discussed in~\cite{Alday:2017gde} where it was pointed out that the gravitational interaction can counter-intuitively induce {\em positive} contributions to anomalous dimensions for values of spin up to $2$. Notice that double-trace operators whose leading coupling is given by~\eqref{eq:c0ksingsing} can be viewed also as affine primaries. Since in each affine block there is an infinite number of quasi-primaries, in the $K3$ theory we have quasi-primaries with the same positive anomalous dimensions as the $k=0,2$ cases in~\eqref{eq:AnomDimSing} and arbitrarily high spin. This is due to the peculiar fact that for 2D CFTs the currents, such as the stress tensor, have twist zero.

It is possible to use the Lorentzian inversion relation~\cite{Caron-Huot:2017vep} to extract a closed form for the $1/N$ terms in the expansion \eqref{eq:n1datacoeff} of coefficients of global blocks with $k>2$ and then we use the explicit data~\eqref{eq:AkBk} to fix the low $k$ values:
\begin{equation} \label{eq:c1ksingsing}
    \begin{aligned}
        \abs{c^{\mathrm{sing},0}_{(1)(0,0)}}^2 &= \frac{1}{2}-\frac{7N_f}{36}\ ,\ \ \abs{c^{\mathrm{sing},1}_{(1)(0,0)}}^2 = 0\ ,\ \  \abs{c^{\mathrm{sing},2}_{(1)(0,0)}}^2 =\frac{1}{9} - \frac{37N_f}{2700}\ ,\\
        \abs{c^{\mathrm{sing},k}_{(1)(0,0)}}^2 &= \frac{1+(-1)^k}{4}\frac{(k!)^2}{(2k)!}\Big(4H_{2k} - 4H_{k}-1\Big) \quad\text{for }\ k>2 \ ,
    \end{aligned}
\end{equation}
where the $H_k=\sum_{n=1}^{k}n^{-1}$ are harmonic numbers. It is straightforward to select the other irreducible flavour representations and here we quote some key results, focusing always on double-trace operators of the class $\mathcal{O}_{\!f_3}\pd^k\bar{\mathcal{O}}_{\!f_4}$. We note that for $(j,\bar j)=(0,0)$ there are no contributions of this type for the antisymmetric representation in the flavour indices. By expanding the R-symmetry singlet and flavour symmetric traceless projection $C^{\mathrm{sym}}_{1\,(0,0)}$ we find the following CFT data
\begin{equation} \label{eq:c0ksingsym}
    \abs{c^{\mathrm{sym},k}_{(0)(0,0)}}^2 =  \Big(1+(-1)^k\Big)\frac{(k!)^2}{4(2k)!} \ \ , \qquad \gamma_{(1)(0,0)}^{\mathrm{sym},k} = -2\quad\forall\,k \text{ even}\ ,
\end{equation}
and
\begin{equation}\label{eq:c1ksingsym}
    \abs{c^{\mathrm{sym},k}_{(1)(0,0)}}^2 = \frac{1+(-1)^k}{4}\frac{(k!)^2}{(2k)!}\Big(4H_{2k} - 4H_{k}-1\Big) + \delta_{k,0} \ .
\end{equation}

As a final example, we provide the anomalous dimensions for another R-symmetry representation, that with $(j,\jb)=(1,0)$. In this case only double-trace operators with odd values of $k$ contribute and in the flavour-singlet sector we have the data
\begin{equation} \label{eq:c0ktripsing}
    \abs{c^{\mathrm{sing},k}_{(0)(1,0)}}^2 = \Big(1+(-1)^{k+1}\Big)\frac{(k!)^2}{4(2k)!} \ \ ,\quad \gamma_{(1)(1,0)}^{\mathrm{sing},k} = \left\{0,\frac{N_f}{3}-2,0,-2,0,-2,\dots\right\} \ ,
\end{equation}
and
\begin{equation}\label{eq:c1ktripsing}
    \abs{c^{\mathrm{sing},k}_{(1)(1,0)}}^2 = \frac{1+(-1)^{k+1}}{4}\frac{(k!)^2}{(2k)!}\Big(4H_{2k} - 4H_{k}-1\Big) -\frac{7N_f}{72}\delta_{k,1} \ .
\end{equation}
Once again these anomalous dimensions are all negative for $N_f=5$, while in the $K3$ case $\gamma_{(1)(1,0)}^{\mathrm{sing},1}\big|_{N_{\!f}=21} = 5$, a shifted spectrum compared with \eqref{eq:AnomDimSing}. Finally, in the flavour-symmetric sector we have the data
\begin{equation}\label{eq:c0ktripsym}
    \abs{c^{\mathrm{sym},k}_{(0)(1,0)}}^2 =  \Big(1+(-1)^{k+1}\Big) \frac{(k!)^2}{4(2k)!}\ \ , \quad \gamma_{(1)(1,0)}^{\mathrm{sym},k} = -2 \ \ \forall\ k\ \text{odd} \ ,
\end{equation}
and
\begin{equation}\label{eq:c1ktripsym}
    \abs{c^{\mathrm{sym},k}_{(1)(1,0)}}^2 = \frac{1+(-1)^{k+1}}{4}\frac{(k!)^2}{(2k)!}\Big(4H_{2k} - 4H_{k}-1\Big) \ .
\end{equation}
One final comment is that the $c_{(1)}$ data matches that computed from the inversion formula in Appendix~\ref{app:inversion} since in both cases what is being extracted is simply the coefficient of global blocks, which as we have noted, does not necessarily give just the OPE coefficients of the double-trace operators $\bar{\mathcal{O}}\pd^k\mathcal{O}$ due to the mixing with triple-traces.

\subsubsection{The $n=2$ case}

While we will not attempt a systematic analysis of the $z,\bar{z}\to 1$ limit for the correlator containing double-traces, we will show in an explicit example how to extract from the correlator~\eqref{eq:c2conn} the OPE coefficients involving three double-trace operators.

A first general feature of the result~\eqref{eq:c2conn} is that there are no terms proportional to $\log^2|1-z|^2$ as $z,\bar{z}\to 1$. From the CFT point of view this is the result of cancellation between different terms entering in~\eqref{eq:n=2ex}. In order to see this let us define, in analogy with the $n=1$ case~\eqref{eq:n1datacoeff}, the couplings appearing in the OPE expansion: for instance for the flavour singlet exchange we have
\begin{equation} \label{eq:C2dt}
    \langle\gro{\bar{\mathcal{O}}^{\,2}_{\!f}}\, \gro{\mathcal{O}^{\,2}_{\!f}}\, (\bar{\mathcal{O}}\mathcal{O})_{k}\rangle\langle(\bar{\mathcal{O}}\mathcal{O})_{k} \mathcal{O}^{(\alpha,\dot\alpha)}_{\!g}\mathcal{O}^{(\beta,\dot\beta)}_{\!g}\rangle \mathcal{P}^{\mathrm{sing}}_{\!ff,\,gg} = M^{\mathrm{sing},k}_{(0)(j,\jb)}\,\sigma^{\alpha\beta}_{(j)}\sigma^{\dot\alpha\dot\beta}_{(\bar j)} + O(N^{-1}) \ ,
\end{equation}
where $M_{(0)}$ mixes information about the $c_{(0)}$ discussed in the previous section and new CFT data $\langle\gro{\bar{\mathcal{O}}^{\,2}_{\!f}}\,\gro{\mathcal{O}^{\,2}_{\!f}}\!(\bar{\mathcal{O}}\mathcal{O})_{k}\rangle$ capturing the couplings between three double-trace operators. As explained above, the subleading coefficients $M_{(1)}$ and $M_{(2)}$ mix the $1/N$ and $1/N^2$ corrections to the couplings in \eqref{eq:C2dt} with new couplings of triple- or quadruple-trace primaries. The leading order $M_{(0)}$ can be easily derived by using the generalised free theory at infinite $N$ which is described by the correlators
\begin{equation} \label{eq:n=2Corrs}
    \begin{aligned}
      \langle :\!\!(\mathcal{O}^{--}_{\!f_1})^2\!\!:\, :\!\!(\mathcal{O}^{++}_{\!f_2})^2\!\!:\mathcal{O}^{++}_{\!f_3}\mathcal{O}^{--}_{\!f_4}\rangle & \approx \frac{1}{\abs{1-z}^2}\Big[\delta_{f_1f_2}\delta_{f_3f_4} + 2\abs{1-z}^2 \delta_{f_1f_3}\delta_{f_2f_4}\Big] \,,\\
      \langle :\!\!(\mathcal{O}^{--}_{\!f_1})^2\!\!:\, :\!\!(\mathcal{O}^{++}_{\!f_2})^2\!\!:\mathcal{O}^{--}_{\!f_3}\mathcal{O}^{++}_{\!f_4}\rangle & \approx \frac{1}{\abs{1-z}^2}\bigg[\delta_{f_1f_2}\delta_{f_3f_4} + \frac{2\abs{1-z}^2}{\abs{z}^2} \delta_{f_1f_4}\delta_{f_2f_3}\bigg] \,,\\
      \langle :\!\!(\mathcal{O}^{--}_{\!f_1})^2\!\!:\, :\!\!(\mathcal{O}^{++}_{\!f_2})^2\!\!:\mathcal{O}^{+-}_{\!f_3}\mathcal{O}^{-+}_{\!f_4}\rangle & \approx -\frac{1}{\abs{1-z}^2}\delta_{f_1f_2}\delta_{f_3f_4} \,,\\
      \langle :\!\!(\mathcal{O}^{--}_{\!f_1})^2\!\!:\, :\!\!(\mathcal{O}^{++}_{\!f_2})^2\!\!:\mathcal{O}^{-+}_{\!f_3}\mathcal{O}^{+-}_{\!f_4}\rangle & \approx -\frac{1}{\abs{1-z}^2}\delta_{f_1f_2}\delta_{f_3f_4} \,,
    \end{aligned}
\end{equation}
where, as usual, we assumed that the double-trace operators are normalised to one. Then it is straightforward to project these results onto the R-symmetry and flavour irreducible representations, as done before, and obtain
\begin{equation}\label{eq:M2c}
  \begin{aligned}
    M^{\mathrm{sing},k}_{(0)(0,0)}= 2 \abs{c^{\mathrm{sing},k}_{(0)(0,0)}}^2 \ , & \quad M^{\mathrm{sing},k}_{(0)(1,0)}= 2\abs{c^{\mathrm{sing},k}_{(0)(1,0)}}^2 \ ,\\  M^{\mathrm{sym},k}_{(0)(0,0)}= 2 \abs{c^{\mathrm{sym},k}_{(0)(0,0)}}^2 \ , & \quad M^{\mathrm{sym},k}_{(0)(1,0)}= 2\abs{c^{\mathrm{sym},k}_{(0)(1,0)}}^2 \ ,
    \end{aligned}
\end{equation}
where the extra factor of $2$ simply follows from the fact that there are twice as many generalised free field diagrams connecting the single- and double-trace operators.

The next contribution to $\gro{C_{2}}$, beyond the one captured by the generalised free theory, is given in~\eqref{eq:c2conn}. Notice that each term in that result is multiplied by a factor of $|z|^2$ as is the case for the terms with $D$-functions in~\eqref{eq:s1fla}. This reflects a particular choice of R-symmetry quantum numbers. We can perform the projection on the R-symmetry singlet exactly as for the $n=1$ correlators and obtain
\begin{align} \label{eq:c2singf1}
    \!\gro{C_{2,(0,0)}} \!&=\! \frac{1+ z +\bar{z} + |z|^2}{N^2}\!\Biggr[\frac{6 i (z+ \zb) (z^2 + 10 z \zb +\zb^2) }{(z-\zb)^5} P_4(z ,\zb)  \! - \! \frac{12 (z^2 + 4 z \zb + \zb^2)}{(z-\zb)^4} P_3(z, \zb)\nonumber \\* 
	& \qquad  + \frac{8 i (z+ \zb)}{(z-\zb)^3}P_2(z, \zb) + \frac{2}{(z-\zb)^2}\, \log |1-z|+  \frac{ ( z + \zb -2 z \zb) }{(z- \zb)^2 |1-z|^2}\log|z|  \nonumber\\
	& \qquad +  \frac{(z+ \zb) }{|1-z|^2 (z-\zb)^2}\big(\log|z|\,\big)^2\Biggr]\,.
\end{align}
We don't have enough data to perform also the decomposition on the flavour irreducible representations since this would require a generalisation of Eq.~\eqref{eq:fullcorr} to the equal flavour case. Thus from now on we focus on the correlator~\eqref{eq:c2singf1} with a pair of fixed (and different) flavours. On general grounds, by focusing on the $(1-z)^0 (1-\bar{z})^0$ terms in the $z,\bar{z}\to 1$ expansion of the connected combination $\gro{C_2}$, we have up to order ${O}(1/N)$
\begin{align} \label{eq:M2conn}
    \gro{C_{2,(0,0)}} &\approx \sum_\mathrm{flav} \mathcal{P}^\mathrm{flav}_{\!ff,\,gg} \left\{ M^{\mathrm{flav},0}_{(0),(0,0)} -2 \abs{c^{\mathrm{flav},0}_{(0),(0,0)}}^2 \right. + \\
    &\ \ \left.\frac{1}{N} \left[ M^{\mathrm{flav},0}_{(1),(0,0)} -2 \abs{c^{\mathrm{flav},0}_{(1),(0,0)}}^2 + \bigg(\!M^{\mathrm{flav},0}_{(0),(0,0)} - 2 \abs{c^{\mathrm{flav},0}_{(0),(0,0)}}^2 \bigg) \gamma^{\mathrm{flav},0}_{(1),(0,0)} \log\abs{1-z}^2\right]\right\} \,, \nonumber
\end{align}
where the sum is over the flavour representations `$\mathrm{flav}$' of the double-trace operator exchanged. Since the correlator under investigation has two pairs of external states with different flavours we have to take $f\neq g$ in the projectors \eqref{eq:projectors}. We know that the $n=2$ connected correlators start at order ${O}(1/N^2)$, so the contributions in~\eqref{eq:M2conn} must cancel. The first line vanishes thanks to the constraints~\eqref{eq:M2c}, following from the generalised free theory and the same is true for the term proportional to $\log|1-z|^2$ in the second line. The absence of a rational term of order $1/N$ yields a relation between the couplings at ${O}(1/N)$
\begin{equation}
  \label{eq:M1c1}
      \left(M_{(1),(0,0)}^{\mathrm{sing},0} - 2 \abs{c_{(1),(0,0)}^{\mathrm{sing},0}}^2\right) - \left(M_{(1),(0,0)}^{\mathrm{sym},0} - 2 \abs{c_{(1),(0,0)}^{\mathrm{sym},0}}^2 \right)= 0 \ .
\end{equation}
We then consider the order ${O}(1/N^2)$ terms in the OPE expansion~\eqref{eq:M2conn}
\begin{align} \label{eq:M2conn2}
    \gro{C_{2,(0,0)}} &\approx \frac1{N^2} \sum_\mathrm{flav} \mathcal{P}^\mathrm{flav}_{\!ff,\,gg} \bigg\{ M^{\mathrm{flav},0}_{(2),(0,0)} - 2 \abs{c^{\mathrm{flav},0}_{(2),(0,0)}}^2 + \bigg[ \bigg( M^{\mathrm{flav},0}_{(0),(0,0)} - 2 \abs{c^{\mathrm{flav},0}_{(0),(0,0)}}^2\bigg) \gamma^{\mathrm{flav},0}_{(2),(0,0)} \nonumber\\
    &\qquad\qquad\qquad\quad+ \bigg( M^{\mathrm{flav},0}_{(1),(0,0)} - 2 \abs{c^{\mathrm{flav},0}_{(1),(0,0)}}^2\bigg) \gamma^{\mathrm{flav},0}_{(1),(0,0)}\bigg] \log\abs{1-z}^2 \nonumber\\
    &\qquad\qquad\qquad\quad+ \frac{1}{2} \bigg( M^{\mathrm{flav},0}_{(0),(0,0)} - 2 \abs{c^{\mathrm{flav},0}_{(0),(0,0)}}^2\bigg) \Big(\gamma^{\mathrm{flav},0}_{(1),(0,0)}\Big)^2 \log^2\abs{1-z}^2\bigg\} \ .
\end{align}
Again, thanks to~\eqref{eq:M2c} the last line vanishes, thus explaining the absence of terms proportional to $\log^2|1-z|^2$ in the OPE expansion of the holographic results. The same pattern holds for the term proportional to $\gamma_{(2)}$ in the first line. Thus the term with $\log|1-z|^2$ is determined by data at order ${O}(1/N)$. The extension of the analysis to exchanged operators of the form \mbox{$(\mathcal{O}\bar{\mathcal{O}})_{k>0}\;\sim \;\mathcal{O}_{\!f_3}\pd^{k>0}\bar{\mathcal{O}}_{\!f_4}$} is complicated by the fact that they can mix, at order in $1/N$, with triple-trace primaries. We leave this analysis for the future.

\section{Summary and outlook}
\label{sec:conclusions}

We have studied correlation functions involving two single-trace and two multi-trace light operators obtained from the light limit of HHLL holographic correlators.
Since these HHLL correlators are computed from quadratic fluctuations around the classical supergravity background dual the heavy operator, it is the connected tree-level Witten diagram contributions to these $n$-trace correlators that are extracted. 
By way of a particular HHLL correlator, whose expression is known exactly in momentum space, we derived explicit expressions for correlation functions involving double-trace operators \eqref{eq:c2conn} (for correlators involving higher multi-trace operators see \eqref{eq:Bcor-short}).
The justification for our method of construction is that the behaviour of these multi-trace correlators in the various OPE channels is consistent in all of the checks we perform, for example, we reproduce the behaviour of the $n=2$ correlator in the $\zb\to1$ light-cone limit -- dominated by the exchange of currents -- from suitable conformal blocks and in the \mbox{$z,\zb\to0$} Euclidean OPE from calculations at the free point involving protected multi-trace operators.

Since 4-point multi-trace correlators can be viewed as a particular kinematical limit of higher-point correlation functions of single-traces, we argued that, for 6-point functions and higher, there are Witten diagrams that cannot be reduced to a finite sum of contact diagrams. Hence, the usual D-functions are not be a sufficient basis of functions for this class of correlators. 
Indeed, from expressions such as \eqref{eq:c2conn}, we see that the
new ingredient appearing  in  correlators with multi-traces are the Bloch-Wigner-Ramakrishnan polylogarithm functions.
These are particular combinations of (higher-order) polylogarithm functions exhibiting simpler analytic properties than the ``bare'' polylogarithms. 

It would be interesting to see whether other explicit computations of higher-point holographic correlators, in any kinematic limit, also have natural descriptions in terms of Bloch-Wigner-Ramakrishnan polylogarithms -- in AdS$_3$ or otherwise. 
We believe that these functions appear generically in holographic correlators involving multi-trace operators.
Furthermore, we have argued that despite such correlators derived from HHLL correlation functions not being exactly equal to those obtained from an OPE limit of higher-point functions, their functional structure is the same.
Thus the conclusions made about the generic appearance of Bloch-Wigner-Ramakrishnan polylogarithm  functions in multi-trace correlation functions apply to higher-order correlators as well.
However, there might exist an even more appropriate basis of functions, possibly with a simple description in Mellin space. This new basis would contain higher-order analogues of the $D_{\Delta_1, \Delta_2, \Delta_3, \Delta_4}$ functions, commonly appearing
in holographic correlators (see for example \cite{D'Hoker:1999ea, D'Hoker:1999jc}) which are just constants when written in Mellin space \cite{Penedones:2010ue}. It would be interesting to find a Mellin description of our multi-trace connected correlators, however, the correlators with a simple Mellin form are likely to be those involving multi-trace operators formed from the OPE limit of single-trace operators, rather than those we have obtained from a HHLL correlator. If such Mellin transforms were to be found, it would be interesting to see whether they are related to the corresponding tree-level Feynman diagrams (see figures~\ref{Fig:WDB4} and~\ref{Fig:2}), with some of the Mellin-variables  set to zero in order to implement the relevant OPE limit. A Mellin space formulation would also provide an explicit avenue towards the flat-space limit where  checks with the relevant amplitudes in Minkowski space~\cite{Heydeman:2018dje,Schwarz:2019aat} could be made. For a summary of the Mellin space description of a wide class of AdS$_3\times S^3$ holographic correlators of single-trace operators see~\cite{Wen:2021lio}.

The known correlators of the $n=1$ type are sufficient to extract dynamical data, such as the anomalous dimensions, of individual R-symmetry and flavour irreducible representations for the family of non-protected double-trace operators of the form \mbox{$\bar{\mathcal{O}}_{\!f}\pd^k\mathcal{O}_{\!g}$} exchanged in the $z,\zb\to1$ OPE channel. 
This unmixed data of such operators (that are degenerate at leading order in large $N$) yields the result that, for the theory on $K3$, the anomalous dimensions for exchanged operators of spin $k\leq2$ are positive.
To the best of our knowledge, this gives the first known example of a theory with such behaviour -- the possibility of which was first discussed in \cite{Alday:2017gde}.
Using the $n=2$ correlators it is possible to derive constraints on the 3-point coupling involving 3 double-trace operators, of the form $\langle\gro{\bar{\mathcal{O}}^{\,2}_{\!f}}\, \gro{\mathcal{O}^{\,2}_{\!f}} \! (\bar{\mathcal{O}}\mathcal{O})_{k=0}\rangle$. For $k>0$, there is non-trivial mixing of these primaries with new triple-trace primaries.
The full unmixing of these operators' data would require the input of a correlator involving two double-trace and two single-trace operators, all with identical flavour index. Such a correlator is not known currently.

Considering the very different qualitative behaviours of the order $1/N$ anomalous dimensions of the double-trace operators $\bar{\mathcal{O}}\pd^k\mathcal{O}$ in the cases of the theory compactified on $T^4$ and $K3$, is it possible that this suggests a fundamental difference between these theories. It would be interesting here to have access to similar LLLL correlators but containing different single-trace operators in order to see whether this pattern in anomalous dimensions is upheld by other families of double-trace operators in these theories.

The smoothness of the limit of correlators between our ``heavy scaling regime" in which the dimension of the multi-trace operators $\mathcal{O}^{\,n}$ scales as $\Delta_{n}\sim c$ and the ``light scaling regime" in which $\Delta_{n}\sim 1$ may be an unexpected feature, but there is now a mounting body of evidence that this is the case~\cite{Giusto:2018ovt,Giusto:2019pxc,Giusto:2020neo}.
The fact that correlation functions of four light operators have been demonstrated to be obtainable from HHLL correlators begs the question of whether also correlators involving operators of intermediate dimensions are smoothly related. Two classes of such operators are the BMN~\cite{Berenstein:2002jq,Gava:2002xb} and other semiclassical states  with dimensions scaling as $\Delta\sim c^{1/4}$ and possibly giant graviton states~\cite{McGreevy:2000cw} with $\Delta\sim c^{1/2}$.

\section*{Acknowledgements}

We would like to thank Alessandro Bombini and Andrea Galliani for collaboration at the early stages of this project. We would like to thank Hongliang Jiang and Congkao Wen for discussions, and Francesco Aprile and Michele Santagata for discussions and also for sharing their work~\cite{Aprile:2021mvq} before publication. This work was supported in part by the Science and Technology Facilities Council (STFC) Consolidated Grant ST/P000754/1 {\it String theory, gauge theory \& duality} and by the MIUR-PRIN contract 2017CC72MK003.
N\v{C} is supported by the ERC Grant 787320 - QBH Structure.

\appendix

\section{Derivation of connected multi-particle correlators}
\label{app:corder}

In this appendix we give details on the derivation of the closed form expressions for the correlators in \eqref{eq:corrs}. 
In addition, we present higher order connected correlation functions that are not discussed in the main text, together with explicit expressions of the Bloch-Wigner-Ramakrishnan polylogarithm functions appearing in them.

The starting point is the HHLL correlation function from Eq.~(3.45) of \cite{Bombini:2017sge}, which for completeness we reproduce here 
\begin{equation} \label{eq:Bfullcorr}
	\gro{C}\!(\tau,\sigma)=\left(1-\frac{B^2}{N}\right) \sum_{k=1}^\infty\sum_{\ell\in\mathbb{Z}} e^{i\ell\sigma} \frac{\exp\left[-i(|\ell| + 2 k)\sqrt{1-\frac{B^2}{N}\left(1- \frac{\ell^2}{(|\ell|+2k)^2}\right)}\,\tau\right]}{\sqrt{1-\frac{B^2}{N}\left(1-\frac{\ell^2}{(|\ell|+2k)^2}\right)} }\,.
\end{equation}
The above expression is written in terms of the dimensionless coordinates ($\tau, \sigma$), related to the coordinates ($t, y$) of the boundary of AdS$_3$ through
\begin{align}
	\tau = \frac{t}{R_y}\ , \quad \sigma = \frac{y}{R_y}\ ,
\end{align}
with $R_y$ being the radius of the spatial circle, so that $y \sim y + 2 \pi R_y$.
Despite being written in terms of the cylinder coordinates $(\tau,\sigma)$, the expression in \eqref{eq:Bfullcorr} represents the correlator on the plane, whose complex coordinates $z$ and $\zb$ are defined by
\begin{align} \label{eq:zzbdef}
	z \equiv e^{i(\tau + \sigma)}\ ,\quad \bar{z} \equiv e^{i(\tau - \sigma)}\ .
\end{align}
In the following we will always work on the Euclidean patch, obtained by the usual Wick rotation $\tau \to -i \tau_e$, or on its analytically continued version with $z$ and $\bar z$ being independent complex coordinates with $\bar z\neq z^*$. Our goal is to rewrite the correlator \eqref{eq:Bfullcorr} in the form of \eqref{eq:serexpC} and so we expand in $B^2/N$
\begin{align} \label{eq:Bserexp1}
	\gro{C}\!(\tau, \sigma) &= \sum_{n=0}^{\infty} \left(\frac{B^2}{N}\right)^{\!n}\, \sum_{\ell \in \mathbb{Z}}\, \sum_{k=1}^{\infty} \, f_n(\tau)\, e^{i \ell \sigma - i (|\ell| + 2 k) \tau}\,,
\end{align}
where we assumed that the sums are well behaved so that their order can be exchanged.
By comparing \eqref{eq:Bserexp1} with \eqref{eq:serexpC}, we 
can extract the connected tree-level correlation functions at each order in $B^2/N$ to get
\begin{align}
	\label{eq:BCnconn}
	\gro{C_{n}}\; = \frac{n!}{N^n}\,\sum_{\ell \in \mathbb{Z}}\, \sum_{k=1}^{\infty} \, f_n(\tau)\, e^{i \ell \sigma - i (|\ell| + 2 k) \tau}\,,
\end{align}
and in what follows we show how to systematically evaluate these double sums.

In \eqref{eq:Bserexp1} we denote by $f_n(\tau)$ polynomial functions of $\tau$ (generically of degree $n$), which are read off by performing the explicit expansion of \eqref{eq:Bfullcorr}. 
The first few are given by
\begin{small}
	\begin{subequations}
		\label{eq:Bfk}
		\begin{align}
			f_0(\tau) &=  1\,,\\*
			f_1(\tau) & =  -\frac{1}{2} - \frac{\sn^2}{2 \tn^2} -i \tau \left(\frac{\sn^2}{2\tn}- \frac{\tn}{2}\right) \label{eq:Bfk1}\,,\\*
			f_2(\tau) & = - \frac{1}{8} + \frac{3 \sn^4}{8\tn^4} - \frac{\sn^2}{4\tn^2} - i \tau \left(- \frac{3 \sn^4}{8\tn^3}+ \frac{\sn^2}{4\tn} + \frac{\tn}{8}\right)  \nonumber\\*
			& \quad + (-i\tau)^2\left(- \frac{\sn^2}{4} + \frac{\sn^4}{8 \tn^2} + \frac{\tn^2}{8} \right)\,, \\
			f_3(\tau) & =  - \frac{1}{16} - \frac{5\sn^6}{16\tn^6} + \frac{9\sn^4}{16\tn^4} - \frac{3\sn^2}{16\tn^2} - i \tau \left(\frac{5\sn^6}{16\tn^5} \right.
			\nonumber  \\*
			& \quad\left.- \frac{9\sn^4}{16\tn^3} + \frac{3\sn^2}{16\tn} + \frac{\tn}{16} \right)+ (-i\tau)^2 \left( -\frac{\sn^2}{8}- \frac{\sn^6}{8\tn^4} \right. \nonumber\\*
			&\quad\left. + \frac{\sn^4}{4\tn^2} \right) +(-i\tau)^3 \left(\frac{\sn^6}{48\tn^3} - \frac{\sn^4}{16\tn} + \frac{\sn^2\tn}{16} - \frac{\tn^3}{48}\right)\,.
		\end{align}
	\end{subequations}
\end{small}
Since $f_0(\tau)$ is trivial, it follows simply that
\begin{subequations}
\begin{align}
	C_{0} &=  \sum_{\ell \in \mathbb{Z}}\, \sum_{k=1}^{\infty} \, e^{i \ell \sigma - i (|\ell| + 2 k) \tau}\label{eq:Bdoublesumc0}
	\\
	&= 	\frac{1}{\left(1- e^{i(\tau + \sigma)}\right)\left(1- e^{i(\tau - \sigma)}\right)}= 
	- \left[ \frac{\plog{0}{e^{-i(\tau + \sigma)}}}{1- e^{-2i\sigma}} + \frac{\plog{0}{e^{-i(\tau -\sigma)}}}{1- e^{2i\sigma}}\right]\,,\label{eq:Bc0}
\end{align}
\end{subequations}
where in the last equality we have used that $\plog{0}{x} = x/(1-x)$.
Finally, by using the complex coordinates given in \eqref{eq:zzbdef} the correlator can be expressed in the simple form
\begin{align}
	\label{eq:Bc0conn}
	C_{0}\!\left(z,\zb\right) = \frac{1}{\left|1-z\right|^2}\ .
\end{align}

To find closed form expressions for higher order correlators we observe that the polynomials $f_n(\tau)$ are generically of the form
\begin{align}
	\label{eq:Bfngen}
	f_n(\tau) = \sum_{p=0}^n\sum_{q=0}^n \frac{a_q}{(2n)!!} (-i\tau)^p\, \ell^{2q} \,(|\ell|+ 2k)^{p-2q}\,,
\end{align}
where $a_q \in \mathbb{Z}$ are some integers.
Since in \eqref{eq:BCnconn}, these functions are multiplied by $e^{i \ell \sigma - i (|\ell| + 2 k) \tau}$, the appropriate powers of $\ell$ and $(|\ell|+ 2k)$ can be obtained term-wise by differentiation or integration of \eqref{eq:Bdoublesumc0} using%
\footnote{Integration over $\sigma$ is not needed since $\ell$ appears only with positive powers in \eqref{eq:Bfngen}.}
\begin{subequations}
	\label{eq:Bderint}
	\begin{align}
		i\partial_\tau 	C_{0} &=  \sum_{\ell \in \mathbb{Z}}\, \sum_{k=1}^{\infty} \,(|\ell|+2k)\, e^{i \ell \sigma - i (|\ell| + 2 k) \tau}\,,\\*
		\frac{1}{i} \int^{\tau} d\tau_1 \, 	C_{0} &= \sum_{\ell \in \mathbb{Z}}\, \sum_{k=1}^{\infty} \,\frac{1}{|\ell|+2k}\, e^{i \ell \sigma - i (|\ell| + 2 k) \tau}\,,\label{eq:Binttau}\\
		- i \partial_\sigma C_{0} &= \sum_{\ell \in \mathbb{Z}}\, \sum_{k=1}^{\infty} \, \ell\,e^{i \ell \sigma - i (|\ell| + 2 k) \tau}\,,
	\end{align}
\end{subequations}
where we have chosen the constant of integration in \eqref{eq:Binttau} to vanish.
Importantly, we assume that the sums on the right-hand side of these expressions are convergent so that term-wise differentiation and integration is well defined. 

In practice, the left-hand side of \eqref{eq:Bderint} is obtained by  using \eqref{eq:Bc0} written in terms of $\rm Li_{0}$ functions, since in that form the $\tau$ variable appears only in the polylogarithm functions.
Thus one can use the recursion relations \eqref{eq:polylogrec} to show that 
\begin{align}\label{eq:BIneq}
	\cI_n^{\,\tau}(\tau, \sigma) & \equiv
	\begin{cases} 
		(i)^n\, \partial_{\tau}^n \,C_{0} \hfill n \geq 0\\
		(i)^n\int^\tau d\tau_1 \int^{\tau_1}d\tau_2 \int\ldots \int^{\tau_{|n|-1}} d\tau_{|n|}\,C_{0} \qquad \  n < 0
	\end{cases}\nonumber\\*
	& = - \left[ \frac{\plog{-n}{e^{-i(\tau + \sigma)}}}{1- e^{-2i\sigma}} + \frac{\plog{-n}{e^{-i(\tau -\sigma)}}}{1- e^{2i\sigma}}\right] \,,
\end{align}
which gives a closed form expression for any integer power of $(|\ell|+ 2k)$ appearing in \eqref{eq:Bfngen}.
Differentiating \eqref{eq:BIneq} with respect to $\sigma$ is trivial, but cumbersome, and we are not aware of any closed form expression for such an action. 
However, since $\tau$ and $\sigma$ are independent one can always perform any $\tau$ operation first, after which the remaining $\sigma$ differentiation is easily performed.
All in all, this allows us to algorithmically translate the expansion polynomials $f_n(\tau)$ into connected tree-level correlators at any order and after rewriting the results in terms of $z$ and $\zb$, we get 
\begin{small}
	\begin{subequations}
		\label{eq:Bcor-short}
		\begin{align}
			C_{0} (z, \zb)&= \frac{1}{|1-z|^2}\,,\\
			C_{1} (z, \zb)&= \frac{1}{N}\biggr[\!- \frac{i }{2} r_2\, P_2(z, \zb)+ \frac12 r_1\! \left( 2\log|1-z| + \frac{z + \zb -2 z \zb}{|1-z|^2} \log|z|\right) - \frac{1}{|1-z|^2}\biggr]\,,
			\\
			\gro{C_{2}}\! (z, \zb)&= \frac{2}{N^2}\biggr[\frac{3 i}{8} r_4 \, P_4 (z, \zb) + \frac32 r_3\, P_3(z, \zb) + 2 i r_2\!\left( P_2 (z, \zb)- \frac{i}{8} \frac{z-\zb}{|1-z|^2} \big(\log|z|\,\big)^2\right)\nonumber\\*
			&\quad\qquad  - \frac12 r_1\!\left( 2\log|1-z| + \frac{z+ \zb - 2 z \zb}{|1-z|^2}\log|z|\right) \biggr]\,,
			\\
			\gro{C_{3}}\! (z, \zb)&=\frac{6}{N^3}\biggr[- \frac{5i}{16} r_6 \! \left( P_6(z, \zb) + \frac{1}{15}\big(\log|z|\,\big)^2 \,P_4(z, \zb)\right)  - \frac{15}{8} r_5\biggr( P_5(z, \zb) \nonumber\\*
			& \qquad \quad \left. + \frac{1}{15}\big(\log|z|\,\big)^2 P_3(z, \zb)\right) - \frac{33 i}{8} r_4\! \left( P_4(z, \zb) + \frac{2}{33}\big(\log|z|\,\big)^2 P_2(z, \zb)\right)  \nonumber \\*
			&\qquad  \quad  - 4 r_3 \!\left( P_3(z, \zb) -\frac{1}{24}\big(\log|z|\,\big)^2\log|1-z|  -\frac{1}{48} \frac{z+\zb- 2 z \zb}{|1-z|^2}\, \big(\log|z|\,\big)^3\right)
			\nonumber \\*
			& \qquad \quad - \frac{3 i }{2 }r_2\! \left( P_2(z, \zb) - \frac{i}{6} \frac{z-\zb}{|1-z|^2}\big(\log|z|\,\big)^2\right)\biggr] \,,\\
			\gro{C_{4}}\! (z, \zb)&= \frac{24}{N^4}\biggr[\frac{35 i }{128} r_8 \!\left(P_8(z, \zb) + \frac{2}{21} \big(\log|z|\,\big)^2\, P_6(z, \zb) \right) + \frac{35}{16} r_7 \biggr(P_7(z, \zb) \nonumber\\*
			&\qquad \quad  \left. + \frac{2}{21} \big(\log|z|\,\big)^2 P_5(z, \zb) \right) +\frac{55i}{8} r_6\!\left(P_6(z, \zb) + \frac{31}{330} \, \big(\log|z|\,\big)^2 P_4(z, \zb) \right)  \nonumber\\*
			& \qquad  \quad  + \frac{85}{8}\,r_5 \!\left(P_5(z, \zb)+ \frac{23}{255}\, \big(\log|z|\,\big)^2\,P_3(z, \zb)\right) + \frac{65i}{8} r_4 \biggr( P_4(z, \zb)\nonumber\\*
			& \qquad\quad  \left.+ \frac{16}{195} \big(\log|z|\,\big)^2 P_2(z, \zb)- \frac{i}{6240} \frac{z-\zb}{|1-z|^2}\, \big(\log|z|\,\big)^4\right) + \frac{5}{2} r_3 \biggr(P_3(z, \zb)
			\nonumber\\*
			& \qquad \quad \left. - \frac{1}{15} \big(\log|z|\,\big)^2 \log|1-z| -  \frac{1}{30} \frac{z+\zb- 2z \,\zb}{|1-z|^2}\big(\log|z|\,\big)^3\right)\biggr]\,,
		\end{align}
	\end{subequations}
\end{small}%
where $r_n$ denote  rational functions of $z$ and $\zb$ defined as
\begin{align}
	\label{eq:Bderrat}
	r_n \equiv \big(z \partial_z - \zb \partial_{\zb}\big)^n\left(\frac{z+ \zb}{z- \zb} \right) \,.
\end{align}
We see that higher-order correlation functions involve higher order generalised Bloch-Wigner-Ramakrishnan polylogarithm functions, as defined in Section~\ref{sec:mtc}.
The explicit forms of $P_2$, $P_3$ and $P_4$ are given in \eqref{eq:Pdef}, while the next few read
\begin{subequations}
	\begin{align}
		P_5(z, \zb) &= \frac12\!\left[ \plog{5}{z} + \plog{5}{\zb} - \log|z|\big( \plog{4}{z} + \plog{4}{\zb}\!\big)\!+ \frac13 \big(\log|z|\,\big)^2 \big(\plog{3}{z} + \plog{3}{\zb}\!\big)\right. \nonumber\\*
		& \left.\quad \qquad  + \frac{2}{45}\big(\log|z|\,\big)^4 \log|1-z| \right]
		\,,\\
		P_6(z, \zb) &=  \frac{1}{2i}\!\left[ \plog{6}{z} -\plog{6}{\zb} - \log|z| \big(\plog{5}{z} - \plog{5}{\zb} \!\big)\! + \frac13 \big(\log|z|\,\big)^2 \big(\plog{4}{z}-\plog{4}{\zb}\!\big)
		\right.\nonumber\\*
		&\left.\quad\qquad - \frac{1}{45} \big(\log|z|\,\big)^4 \big(\plog{2}{z} -\plog{2}{\zb}\!\big) \right]
		\,,\\
		P_7(z, \zb) &=  \frac12 \!\left[ \plog{7}{z} + \plog{7}{\zb} - \log|z|\big(\plog{6}{z} + \plog{6}{\zb}\!\big)\!+ \frac13 \big(\log|z|\,\big)^2 \,\big(\plog{5}{z} + \plog{5}{\zb}\!\big)
		\right.\nonumber\\*
		& \left.\quad\qquad   - \frac{1}{45}\big(\log|z|\,\big)^4 \big(\plog{3}{z} + \plog{3}{\zb}\!\big) - \frac{4}{945}\big(\log|z|\,\big)^6\log|1-z|\right] 
		\,,\\
		P_8(z, \zb) &=  \frac{1}{2i}\!\left[\plog{8}{z} -\plog{8}{\zb} - \log|z| \big(\plog{7}{z} - \plog{7}{\zb}\! \big) \!+ \frac13 \big(\log|z|\,\big)^2 \big(\plog{6}{z}-\plog{6}{\zb}\!\big)
		\right.\nonumber\\*
		&\left.\quad\qquad - \frac{1}{45} \big(\log|z|\,\big)^4 \big(\plog{4}{z} -\plog{4}{\zb}\!\big) + \frac{2}{945} \big(\log|z|\,\big)^6 \big(\plog{2}{z} -\plog{2}{\zb}\!\big)\right]
		\,.
	\end{align}
\end{subequations}
We note that when fully written out the expressions \eqref{eq:Bcor-short} match \eqref{eq:corrs}.

\section{Summary of the $n=1$ correlators}
\label{app:n1summ}

\begin{figure}[t]
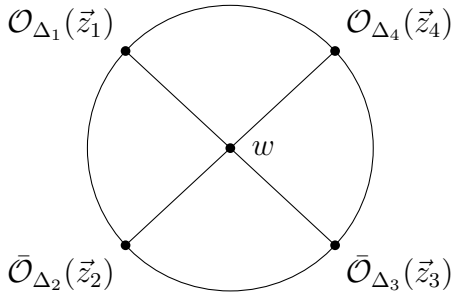

	\begin{center}
		\[\begin{wittendiagram}
			\draw (1.3928,1.29233) node[vertex] node[anchor=south west]{$\mathcal{O}_{\Delta_4}(\vec{z}_4)$};
			\draw (1.3928,-1.29233) node[vertex]node[anchor=north west]{$\bar{\mathcal{O}}_{\Delta_3}(\vec{z}_3)$};
			\draw (1.3928,1.29233) -- (0,0);
			\draw (1.3928,-1.29233) -- (0,0);
			\draw (0,0) node[vertex] node[anchor =west] {$\, \,w$};
			\draw  (-1.3928,1.29233) node[vertex]node[anchor=south east]{$\mathcal{O}_{\Delta_{1}}(\vec{z}_1)$};
			\draw  (-1.3928,-1.29233) node[vertex]node[anchor=north east]{$\bar{\mathcal{O}}_{\Delta_{2}} (\vec{z}_2) $};
			\draw (-1.3928,1.29233) -- (0,0);
			\draw (-1.3928,-1.29233) -- (0,0);
		\end{wittendiagram}\]
	\end{center}
	\caption{The contact Witten diagram corresponding to the integral \eqref{eq:Dfun1}. 
		Four scalar operator insertions $\cO_{\Delta_i}$ with scaling dimensions $\Delta_i$ interact via a quartic vertex in the bulk. 
		In the expression for $D_{\Delta_1, \Delta_2, \Delta_3, \Delta_4}$ one has to integrate over all possible interaction points $w$.
	}
	\label{fig:contact}
\end{figure}

To order $1/N$ all 4-point correlations functions of matter fields $s_1$ in AdS$_3\times S^3$ are known \cite{Rastelli:2019gtj, Giusto:2018ovt}. 
Using the notation 
\begin{equation}
	\label{eq:BCabfi}
	\mathcal{C}^{\alpha\dot{\alpha},\beta\dot{\beta}}_{1,\,f_1 f_2 f_3 f_4}\equiv \langle \mathcal{O}^{--}_{\!f_1}(0)  \mathcal{O}^{++}_{\!f_2}(\infty) \mathcal{O}^{\alpha\dot\alpha}_{\!f_3}(1) \mathcal{O}^{\beta\dot\beta}_{\!f_4}(z,\bar z) \rangle\,,
\end{equation}
one finds the following expressions
\begin{subequations}
	\label{eq:n1corrs}
	\begin{align}
		\label{eq:s1fla}
		\mathcal{C}^{++--}_{1,\,f_1 f_2 f_3 f_4} &=
		\frac{1}{|1-z|^2}\Bigg[\left(1-\frac{1}{N}\right)\Big(\delta_{f_1 f_2} \delta_{f_3 f_4} + |1-z|^2 \delta_{f_1 f_3} \delta_{f_2 f_4}\Big) + 
		\frac{2}{N \pi} |1-z|^2 |z|^2 \,\times \nonumber \\*
		& \quad \left(\delta_{f_1 f_2} \delta_{f_3 f_4} \hat{D}_{1122}(z, \zb) + \delta_{f_1 f_4} \delta_{f_2 f_3} \hat{D}_{2112}(z, \zb) + \delta_{f_1 f_3} \delta_{f_2 f_4} \hat{D}_{1212} (z, \zb)
		\right)\Bigg]\;,\\
		\label{eq:s1flb}
		\mathcal{C}^{--++}_{1,\,f_1 f_2 f_3 f_4} &=
		\frac{1}{|1-z|^2}\Bigg[\left(1-\frac{1}{N}\right)\left(\delta_{f_1 f_2} \delta_{f_3 f_4} + \frac{|1-z|^2}{|z|^2}\delta_{f_1 f_4} \delta_{f_2 f_3} \right) + 
		\frac{2}{N \pi} |1-z|^2 \,\times \nonumber\\*
		& 
		\quad \left(\delta_{f_1 f_2} \delta_{f_3 f_4} \hat{D}_{1122}(z, \zb) + \delta_{f_1 f_4} \delta_{f_2 f_3} \hat{D}_{2112}(z, \zb) + \delta_{f_1 f_3} \delta_{f_2 f_4} \hat{D}_{1212}(z, \zb) \right)\Bigg]\;, \\ 
		\label{eq:s1flc}
		\mathcal{C}^{+--+}_{1,\,f_1 f_2 f_3 f_4} &=
		-\frac{1}{|1-z|^2}\Bigg[\left(1-\frac{1}{N}\right)\delta_{f_1 f_2} \delta_{f_3 f_4} + 
		\frac{2}{N \pi} |1-z|^2 z\Bigr(\delta_{f_1 f_2} \delta_{f_3 f_4} \hat{D}_{1122}(z, \zb)  \nonumber \\*
		&\quad + \delta_{f_1 f_4} \delta_{f_2 f_3} \hat{D}_{2112}(z, \zb) + \delta_{f_1 f_3} \delta_{f_2 f_4} \hat{D}_{1212}(z, \zb) \Bigr)\Bigg]\;,\\
		\label{eq:s1fld}
		\mathcal{C}^{-++-}_{1,\,f_1 f_2 f_3 f_4} &=
		-\frac{1}{|1-z|^2}\Bigg[\left(1-\frac{1}{N}\right)\delta_{f_1 f_2} \delta_{f_3 f_4} + 
		\frac{2}{N \pi} |1-z|^2 \bar{z}\Bigr(\delta_{f_1 f_2} \delta_{f_3 f_4} \hat{D}_{1122}(z, \zb)\nonumber  \\*
		&\quad  + \delta_{f_1 f_4} \delta_{f_2 f_3} \hat{D}_{2112}(z, \zb) + \delta_{f_1 f_3} \delta_{f_2 f_4} \hat{D}_{1212}(z, \zb) \Bigr)\Bigg]\;.
	\end{align}
\end{subequations}
In the above we use the compact notation \mbox{$\hat D_{\Delta_1, \Delta_2, \Delta_3, \Delta_4}$},  which are specific combinations of logarithms of $z$ and $\zb$, and $P_2(z,\zb)$ --  the second order Bloch-Wigner-Ramakrisnan function%
\footnote{Note that while in the  context of this paper it might have been more natural to denote $\hat D_{\Delta_1, \Delta_2, \Delta_3, \Delta_4}$ 
as $\hat P^{(2)}_{\Delta_1, \Delta_2, \Delta_3, \Delta_4}$, we retain the notation commonly used in the literature as generalisations involving higher order BWR functions $P_n$ are not known.}.
These functions naturally arise in four-point correlation functions in the context of the  AdS/CFT correspondence \cite{Maldacena:1997re, Gubser:1998bc, Witten:1998qj}.  
For example, contact Witten diagrams in AdS$_{d+1}$ involving four operators with scaling dimensions $\Delta_{i}$, dual to scalar fields in the bulk, can be written as an integral over four scalar bulk-to-boundary propagators \cite{Dolan:2000ut, Arutyunov:2002fh}
\begin{align} \label{eq:Dfun1}
	D_{\Delta_1, \Delta_2, \Delta_3, \Delta_4}(\vec z_i) =\int d^{d+1}{w}\, \sqrt{g\,} \,\prod_{i=1}^4 K_{\Delta_i}(w;\vec{z}_i)\ ,
\end{align}
where we work in the Poincar\'e patch of AdS$_{d+1}$ in Euclidean signature
\begin{align}
	ds^2 = \frac1{\omega_0^2}\left({dw_0^2 +  d\vec {w}^{\,2}}\right)\,,
\end{align} 
with the flat $d$-dimensional boundary being located at $w_0 = 0$ and $\vec{z}_i$ with $i = 1,2,3,4$ denoting four insertion points of the external operators on this boundary (see figure~\ref{fig:contact}).
In this spacetime, bulk-to-boundary propagators take the form 
\begin{align}
	K_{\Delta_i}(w;\vec{z}_i) = \left[ \frac{w_0}{w_0^2 + (\vec w - \vec z_i)^2}\right]^{\Delta_i}\,,
\end{align}
which, after introducing four Schwinger parameters $t_i$, allows us to rewrite \eqref{eq:Dfun1} as
\begin{align}
	\label{eq:Dfun2}
	D_{\Delta_1, \Delta_2, \Delta_3, \Delta_4}(\vec z_i) =\Gamma\!\left(\frac{\hat \Delta-d}{2}\right) \int_0^\infty \prod_{i=1}^4\bigg[dt_i \frac{t_i^{\Delta_i-1}}{\Gamma(\Delta_i)}\bigg]\, \frac{\pi^{d/2}}{2 T^{\frac{\hat \Delta}{2}}}\,e^{-\sum_{i,j=1}^4 |\vec{z}_{ij}|^2\,\frac{t_i t_j}{2 T}}\,,
\end{align}
where 
\begin{align}
	\vec{z}_{ij} = \vec{z}_i - \vec{z}_j\ , \qquad \hat \Delta = \sum_{i=1}^4 \Delta_i\ , \qquad T = \sum_{i=1}^4 t_i\ .
\end{align}
By rewriting the integral in this form, one notices that differentiating \mbox{$D_{\Delta_1, \Delta_2, \Delta_3, \Delta_4}$}
with respect to $|\vec{z}_{ij}|^2$ one obtains an expression which is proportional to the $D$-function related to contact diagrams with operator insertions that have higher scaling dimensions, for example
\begin{align}
	\label{eq:DerVal}
	\frac{\partial}{\pd |\vec{z}_{12}|^2} D_{\Delta_1, \Delta_2, \Delta_3, \Delta_4}(\vec z_i) = - \frac{2\Delta_1 \Delta_2}{\hat \Delta - d}\, D_{\Delta_1+1, \Delta_2+1, \Delta_3, \Delta_4}\,.
\end{align}
Such relations become especially valuable, since in $d=2$ one can evaluate
\begin{align}
	\label{eq:D1111}
	D_{1111}(\vec{z}_i) = \frac{2\pi i}{ |\vec{z}_{13}|^2 |\vec{z}_{24}|^2 (z-\zb)} \, P_2(z,\zb)\,,
\end{align}
where we used the conformal cross-ratios \eqref{eq:zzbdef} and the second order Bloch-Wigner-Ramakrishnan polylogarithm function \eqref{eq:BWdef1}.
Furthermore, recall that in our analysis we pick a specific gauge \eqref{eq:zgauge} and thus it is convenient to define a new set of
functions
\begin{align}
	\label{eq:DhatDef}
	\hat D_{\Delta_1, \Delta_2, \Delta_3, \Delta_4}(z, \zb) \equiv \lim_{z_2 \to \infty} |z_2|^{2\Delta_2} \, D_{\Delta_1, \Delta_2, \Delta_3, \Delta_4}(z_1 = 0, z_2, z_3 = 1, z_4 = z)\,,
\end{align}
where $z_i$, which denote points on the two-dimensional boundary, are now  complex variables%
\footnote{Note that $\hat D$-functions are  related to $\bar D$-functions, which are often used in the literature \cite{Dolan:2000ut, Arutyunov:2002fh, Penedones:2010ue}, by
	\begin{equation}
		\hat D_{\Delta_1,\Delta_2,\Delta_3,\Delta_4}(z,\bar z) = \frac{\pi\,\Gamma\!\left(\frac{\hat \Delta-2}{2}\right)}{2 \prod_{i=1}^4 \Gamma(\Delta_i)}\,|z|^{\hat \Delta - 2\Delta_1 - 2\Delta_4} |1-z|^{\hat \Delta -2 \Delta_3-2\Delta_4}\bar D_{\Delta_1,\Delta_2,\Delta_3,\Delta_4}(z,\bar z)\,.
\end{equation}}. Using this definition it follows that 
\begin{align}
	\hat D_{1111}(z, \zb) =  \frac{2 \pi i }{z - \zb} \, P_2(z, \zb)\,,
\end{align}
while functions with higher values of the indices can be obtained from $\hat D_{1111}$ using the derivative relations \eqref{eq:DerVal} together with several identities that $\hat D$-functions satisfy, such as%
\footnote{Other identities can be found for example in \cite{Giusto:2018ovt, Bombini:2019vnc}.}
\begin{align} \label{eq:DhatInv}
	\hat D_{\Delta_2, \Delta_1, \Delta_3, \Delta_4} \left( \frac{1}{z}, \frac{1}{\zb}\right) &= |z|^{2 \Delta_4}\, \hat D_{\Delta_1, \Delta_2, \Delta_3, \Delta_4} (z, \zb)\,.
\end{align}
Concretely,  the functions appearing in \eqref{eq:n1corrs} can be written explicitly as
\begin{subequations}
	\label{eq:D12s}
	\begin{align}
		\hat D_{1122}(z, \zb) & = -\frac{2\pi i}{(z-\bar z)^2}\!\left[ \frac{z+\bar z}{z-\bar z} P_2(z,\bar z)+\frac{\log|1-z|^2}{2i}+ \frac{z+\bar z-2 \abs{z}^2}{4i \,|1-z|^2}\log|z|^2\right]\,,
		\\
		\hat D_{2112}(z, \zb) & = -\frac{2\pi i}{(z-\bar z)^2}\!\left[ \frac{2-z-\bar z}{z-\bar z} P_2(z,\bar z)+ \frac{z+\bar z-2\abs{z}^2}{4i \,|z|^2}\log|1-z|^2+\frac{\log|z|^2}{2i}\right]\,,
		\\
		\hat D_{1212}(z, \zb) & = -\frac{2\pi i}{(z-\bar z)^2}\!\left[ \frac{2\abs{z}^2 -z-\bar z}{z-\bar z} P_2(z,\bar z)+ \frac{z+\bar z-2 }{4i }\log|1-z|^2-\frac{z + \zb}{2i}\log|z|^2\right]\,.
	\end{align}
\end{subequations}

Let us conclude with the observation that the $\hat D$-functions written above  can be written schematically as
\begin{align}
	\hat D \sim f_1(z, \zb) P_2(z, \zb) + f_2(z, \zb) \, \log|1-z|^2 + f_3(z, \zb) \, \log|z|^2\,,
\end{align}
where $f_i(z, \zb)$ are some meromorphic functions of $z$ and $\zb$. 
The same structure can also be seen in $\hat D$-functions with higher values for the indices and  generalisations of \eqref{eq:Dfun1} corresponding to higher $n$-point contact diagrams, discussed for example in \cite{Goncalves:2019znr}. 
Since  $\hat D$-functions form the main building blocks of 4-point correlation functions in AdS$_3\times S^3$, it follows that correlators themselves should have the same structure.

\section{Double-trace data from the inversion formula}
\label{app:inversion}

In this section we  derive  the anomalous dimensions and OPE coefficients that were presented in section~\ref{sssec:n1} using the Lorentzian inversion formula \cite{Caron-Huot:2017vep, Simmons-Duffin:2017nub}.
In particular, we are interested in the OPE data of double-trace operators  exchanged in the $z, \zb \to 1$ (or equivalently $z_1 \to z_2$) channel of the $n=1$ correlator in \eqref{eq:Cabfi}, which can be extracted from the singularities of the remaining two channels.

Begin by writing the four-point correlation function as
\begin{equation} \label{eq:CCabfi}
	\mathcal{C}^{\alpha\dot{\alpha},\,\beta\dot{\beta}}_{n=1, f_1 f_2 f_3 f_4} = \langle \mathcal{O}^{--}_{\!f_1}(0)  \mathcal{O}^{++}_{\!f_2}(\infty) \mathcal{O}^{\alpha\dot\alpha}_{\!f_3}(1) \mathcal{O}^{\beta\dot\beta}_{\!f_4}(z,\bar z) \rangle = \frac{1}{|1-z|^2}\cG^{\alpha\dot{\alpha},\,\beta\dot{\beta}}_{n=1,f_1 f_2 f_3 f_4}\left(z, \zb\right)\ ,
\end{equation}
where we used the conformal cross-ratios $z,\zb$ defined in \eqref{eq:crossratio}. 
Here we find it convenient instead to define 
\begin{align}\label{eq:zCRt}
	Z=\frac{z_{12} z_{34}}{z_{13} z_{24}}\quad,\quad \Zb=\frac{\zb_{12} \zb_{34}}{\zb_{13} \zb_{24}}\,,
\end{align}
which are related  to the $z, \zb$ by
\begin{align}
	\label{eq:ztrans}
	z = 1-Z\ , \quad \zb = 1- \Zb\ ,
\end{align}
and as before $z_{ij}= z_i-z_j$.
The use of  these capitalised conformal cross-ratios allows us to make close contact with \cite{Caron-Huot:2017vep} (see also \cite{Alday:2017vkk, Kraus:2018zrn}).
The $s$-channel OPE limit $z_1 \to z_2$, which corresponds to taking $Z, \Zb\to 0$, can be written as a sum over the exchange of quasi-primary operators, with spin $k=\abs{h-\hb}$ and scaling dimension $\Delta=h+\hb$.
The Lorentzian inversion formula states that for large enough spin the CFT data in the $s$-channel is completely encapsulated by a function%
\footnote{Here we suppress flavour and R-symmetry indices until we consider specific correlators.}
\begin{align}\label{eq:Cfun}
	c(h,\hb) \equiv c^t(h,\hb) + (-1)^{k} c^u(h,\hb)\ ,
\end{align}
which is analytic in spin and built from the information of the $t$-channel ($Z\to1$) and $u$-channel ($Z\to\infty$) of the correlator, with
the details of the function depending on the external operators and the dimension of the spacetime.
Assuming $h \geq \hb$, so that $k= h - \hb$ and
\begin{align}
	\label{eq:hhbdef}
	h = \frac{\Delta+k}{2}\,, \qquad \hb = \frac{\Delta-k}{2}\,,
\end{align}
in $d=2$ and for the correlator \eqref{eq:CCabfi}, we use
\begin{align}\label{eq:CtFunhhb}
	c^t(h,\hb) &\equiv \frac{\kappa}{2}\int_0^1 \frac{dZ}{Z^2}\, \frac{d\Zb}{\Zb^2}\, g_h(Z) \,g_{1-\hb}(\Zb)\, \dDisc\!\!\left[\mathcal{G}(Z,\Zb)\right]\,,
\end{align}
where 
\begin{subequations}
	\begin{align}
		\kappa &= \frac{ \Gamma^4\!\left(h\right)}{2 \pi^2\, \Gamma\!\left(2h -1\right) \Gamma\!\left(2h\right)}\,,\\
		g_h(Z) &=  z^h\, {}_2F_{1}(h, h, 2h, Z)\,.
	\end{align}
\end{subequations}
The double discontinuity that picks out the relevant singularities we take to be
\begin{align}\label{eq:dDiscDef}
	\dDisc\!\!\left[\mathcal{G}(Z,\Zb)\right] \equiv \mathcal{G}(Z,\Zb) - \frac12\Big( \mathcal{G}_{\circlearrowright}(Z,\Zb)+ \mathcal{G}_{\circlearrowleft}(Z,\Zb)\Big) \,,
\end{align}
where we analytically continue around $Z=1$, while leaving $\Zb$ fixed as%
\footnote{Note that in \cite{Caron-Huot:2017vep} the analytic continuation is performed around $\Zb=1$, however, there $h$ and $\hb$ are exchanged as compared to \eqref{eq:hhbdef}.}
\begin{subequations}
	\label{eq:AnalCont}
	\begin{align}
		&\mathcal{G}_{\circlearrowright}(Z,\Zb)\,:\qquad (1-Z)\to e^{-2\pi i } (1-Z)\,,\\
		&\mathcal{G}_{\circlearrowleft}(Z,\Zb)\,:\qquad (1-Z)\to e^{2\pi i } (1-Z)\,.
	\end{align}
\end{subequations}
The function $c^{u}(h,\hb)$ can be obtained in a similar manner, only that the analytic continuation is performed around $Z \to \infty$. 
In practice, this can be done by swapping $z_1 \leftrightarrow z_2$, so that $z\to z^{-1}$ and $\zb \to \zb^{-1}$ followed by using \eqref{eq:ztrans}, in which case one can again analytically continue around $Z = 1$ as in \eqref{eq:AnalCont}.

The CFT data is contained in the analytic structure of the inversion function. The location of simple poles gives information about the twists of the exchanged quasi-primary operators, while the residue at that pole is related to the coefficients of \emph{global} blocks in the s-channel. If only one quasi-primary for given conformal dimensions contributes to the correlator under question, these coefficients will simply be the square of the OPE coefficients of external operators with the quasi-primary. 
The precise form of \eqref{eq:Cfun} near a pole corresponding to the exchange of a quasi-primary with left- and right-moving conformal dimensions
$(h_p, \hb_p)$ is given by
\begin{equation} \label{eq:hhbPoles}
	c(h,\hb) \sim - \frac12 \, \frac{C^{\,2}_{h_p,\hb_p}}{\hb- \hb_p}\,,
\end{equation}
where the  factor of $1/2$ comes from replacing the twist $\Delta- k$ with $\hb$. 

We aim to reproduce the CFT data of double-trace operators exchanged in the \mbox{$s$-channel} of \eqref{eq:CCabfi}, having the schematic form 
\mbox{$(\mathcal{O}\bar{\mathcal{O}})_{\nb,k}\;\sim\mathcal{O}_{\!f_3}\pd^{\nb+ k}\pdb^{\nb}\bar{\mathcal{O}}_{\!f_4}$}, for which at large $N$
\begin{subequations} \label{eq:CCFTdata}
	\begin{align}
		h_{\nb,k} &= 1 + \nb + k +  \frac{\gamma_{(1)}^{\nb, k}}{N}+ O(1/N^2) \,,\\*
		\hb_{\nb,k} &= 1 + \nb  + \frac{\gamma_{(1)}^{\nb, k}}{N}+ O(1/N^2) \,,
	\end{align}
\end{subequations}
with ${\gamma_{(1)}^{\nb,k}}$ denoting the anomalous dimensions. We also expand the residues in large $N$ (see Section~\ref{sssec:n1} for a more detailed discussion on the relation of these coefficients to the 3-point functions of the double-trace operators $(\mathcal{O}\bar{\mathcal{O}})_{\nb,k}$)
\begin{equation} \label{eq:coeffexpc}
    C^{\,2}_{h_p\hb_p} = \big| c_{(0)}^{\nb,k}\big|^2 + \frac{1}{N} \big|c_{(1)}^{\nb,k}\big|^2 + O(1/N^2)\,.
\end{equation}
Inserting these expressions into \eqref{eq:hhbPoles} and expanding in $1/N$ yields 
\begin{align} \label{eq:hhbPoles2}
	c(h,\hb) \approx - \frac{1}{2} \left[ \frac{|c_{(0)}^{\nb, k}|^2}{\hb - (1+ \nb)} + \frac{1}{N}\left(\frac{|c_{(1)}^{\nb, k}|^2}{\hb-(1+ \nb)} + \frac{|c_{(0)}^{\nb,k}|^2\, \gamma_{(1)}^{\nb, k}}{\big(\hb-(1+ \nb)\big)^2} \right)\right]+ O(1/N^2)\,,
\end{align}
and thus one is able to extract the unknown quantities by analysing poles of different degrees at $\hb = 1+ \nb$, order by order in  the large $N$ expansion.
In particular, to make contact with the analysis of section~\ref{sssec:n1}, we focus on the minimal-twist operators%
\footnote{To avoid the cluttering of indices and to have notation consistent with section~\ref{sec:lchecks}, the CFT data of double-traces with $\nb =0$ will only have a single index $k$ denoting the spin of the exchanged operator, for example $\gamma_{(1)}^{k} \equiv \gamma_{(1)}^{\nb=0, k}$.}
\mbox{$(\mathcal{O}\bar{\mathcal{O}})_{k}\;\sim\mathcal{O}_{\!f_3}\pd^k\bar{\mathcal{O}}_{\!f_4}$}, with $\nb =0$, for which we can extract the CFT data by other independent methods as well.

Let us now consider the specific example of the $n=1$ correlator \eqref{eq:Cabfi} in the R-symmetry singlet projection
\begin{equation} \label{eq:G00def}
	\begin{aligned}
		&\mathcal{G}_{1,\,f_1 f_2 f_3 f_4}^{(0,0)}(z,\zb) = 
		\left(1-\frac{1}{N}\right)\!\left(\delta_{f_1 f_2} \delta_{f_3 f_4}+ \frac{|1-z|^2}{4} \delta_{f_1 f_3} \delta_{f_2 f_4} + \frac{|1-z|^2}{4 |z|^2}\delta_{f_1 f_4} \delta_{f_2 f_3} \right) \\*
		&
		\ \ \,+ 
		\frac{|1-z|^2}{N \pi} \frac{\abs{1+z}^2}{2} \left(\delta_{f_1 f_2} \delta_{f_3 f_4} \hat{D}_{1122}(z,\zb) + \delta_{f_1 f_4} \delta_{f_2 f_3} \hat{D}_{2112}(z,\zb)  + \delta_{f_1 f_3} \delta_{f_2 f_4} \hat{D}_{1212}(z,\zb)  \right)\,.
	\end{aligned}
\end{equation}
We first analyse the $t$-channel contribution $c^t(h, \hb)$ to the inversion formula. 
Begin by  using \eqref{eq:ztrans} to rewrite $\mathcal{G}_{1,\,f_1 f_2 f_3 f_4}^{(0,0)}$ in terms of $Z$ and $\Zb$, expand the expression in%
\footnote{The choice of this particular combination of $Z$ is convenient as it allows us to evaluate the integrals that appear in the inversion function in a closed form \cite{Alday:2017vkk, Caron-Huot:2017vep}.}
 $(1-Z)/Z$ to find
\begin{align}
	\label{eq:G00tExp}
	&\mathcal{G}_{1,\,f_1 f_2 f_3 f_4}^{(0,0)}(1-Z,1-\Zb)\sim  \frac{\delta_{f_1f_4} \delta_{f_2 f_3}}{4}  \,\widetilde{\cG}_{1}^{(0,0)}(\Zb) \, \frac{Z}{1-Z} + \ldots\,,
\end{align}
where 
\begin{align}
	\label{eq:G00fun}
	\widetilde{\cG}_{1}^{(0,0)}(\Zb) = \left[ \frac{\Zb}{1-\Zb} - \frac{1}{N}\left(  \frac{\Zb}{1-\Zb} + \left(2\frac{\Zb}{1-\Zb} + \frac{\Zb^2}{(1-\Zb)^2}\right) \log\Zb\right)\right]\,.
\end{align}
In \eqref{eq:G00tExp} we only show terms that are non-vanishing after taking the double discontinuity \eqref{eq:dDiscDef}, which are poles and \emph{double} logarithms at $Z = 1$. 
After inserting \eqref{eq:G00tExp} into \eqref{eq:CtFunhhb} the integrals factorise to give
\begin{align} \label{eq:ct}
	c^t(h, \hb) &= \frac{\delta_{f_1f_4}\delta_{f_2f_3}}{4}\, \frac{\kappa}{2}
	\int_0^1\frac{dZ}{Z^2}\, g_h(Z)\, \dDisc\!\!\left[\frac{Z}{1-Z} \right]\,
	\int_0^1\frac{d\Zb}{\Zb^2} \,g_{1-\hb}(\Zb) \, \widetilde{\cG}_{1}^{(0,0)}(\Zb) \,.
\end{align}

The integral involving the double discontinuity has to be evaluated with extra care. Begin by noting that for a generic exponent $p$ \cite{Caron-Huot:2017vep, Alday:2017vkk}
\begin{align} \label{eq:dDiscp}
	\dDisc\!\!\left[\left(\frac{1-Z}{Z}\right)^{\!p}\,\right] = 2 \sin^2(p\pi) \left(\frac{1-Z}{Z}\right)^{\!p}\,,
\end{align}
which would na{\"i}vely vanish for integer $p$, however, the resulting double root precisely cancels out a double pole arising from the integral over $Z$.
Let us rewrite \eqref{eq:ct} as 
\begin{align} \label{eq:ct1}
	c^t(h, \hb) &= \frac{\delta_{f_1f_4}\delta_{f_2f_3}}{4}\, \frac{\kappa}{2}
	\lim_{p\to -1}\int_0^1\frac{dZ}{Z^2}\, g_h(Z)\, \dDisc\!\!\left[\left(\frac{1-Z}{Z}\right)^{\!p\,} \right]\,
	\int_0^1\frac{d\Zb}{\Zb^2} \,g_{1-\hb}(\Zb) \,\widetilde{\cG}_{1}^{(0,0)}(\Zb)\nonumber\\*
	& = \frac{\delta_{f_1f_4}\delta_{f_2f_3}}{4}\, \frac{\kappa}{2}
	\lim_{p\to -1}\left[2\sin^2(p\pi)\,\cI^{p}(h)\right]\times \\*
	&\hspace{2.5cm}\left[ \cI^{-1}(1-\hb) - \frac{1}{N} \Big(\cI^{-1}(1-\hb) + 2\mathcal{J}^{-1}(1-\hb) + \mathcal{J}^{-2}(1-\hb)\Big)\right]\,,\nonumber
\end{align}
where we defined
\begin{subequations} \label{eq:IntDef}
	\begin{align}
		\cI^a(h) &\equiv \int_0^1 \frac{dZ}{Z^2} \, g_h(Z)\,\left( \frac{1-Z}{Z}\right)^{\!a} =\frac{\Gamma(2h)\,\Gamma^2(a+1)\, \Gamma(h-a-1)}{\Gamma^2(h)\,\Gamma(h+a+1)}\,,\label{eq:IntDef1}
		\\
		\mathcal{J}^a(h) &\equiv \int_0^1 \frac{d\Zb}{\Zb^2} \, g_h(\Zb)\,\left( \frac{1-\Zb}{\Zb}\right)^{\!a}\, \log{\Zb}\,.\label{eq:IntDef2}
	\end{align}
\end{subequations}
Integral \eqref{eq:IntDef1}  is evaluated following \cite{Alday:2017vkk} and using this closed from expression we get
\begin{align}
	2\sin^2(p\pi)\cI^p(h) = 2\pi^2\, \frac{\Gamma(2h)\, \Gamma(h-p-1)}{\Gamma^2(h)\,\Gamma(h+p+1)\,\Gamma^2(-p)}\,,
\end{align}
which vanishes only for non-negative integer values of $p$.
It also follows that
\begin{align} \label{eq:cIh}
    \kappa \lim_{p\to -1} \left[2\sin^2(p\pi)\,\cI^{p}(h)\right] = \frac{\Gamma^2(h)}{\Gamma(2h-1)} \,.
\end{align}

Next, we can  repeat this procedure for the $u$-channel of \eqref{eq:G00def}.
As already discussed, we swap $z_1 \leftrightarrow z_2$ causing $z \to z^{-1}$ and $\zb \to \zb^{-1}$, followed by the change to $Z$, $\Zb$ variables using \eqref{eq:ztrans}.
Extracting the singular terms at $Z = 1$ yields 
\begin{align}
	\label{eq:G00uExp}
	&\mathcal{G}_{1,\,f_1 f_2 f_3 f_4}^{(0,0)}\left(\frac{1}{1-Z},\frac{1}{1-\Zb}\right)\sim  \frac{\delta_{f_1f_3} \delta_{f_2 f_4}}{4}  \,\widetilde{\cG}_{1}^{(0,0)}(\Zb) \, \frac{Z}{1-Z} + \ldots\,,
\end{align}
with the function of $\Zb$ again being given by  \eqref{eq:G00fun}%
\footnote{This can be understood from the fact that sending $z \to z^{-1}$ and $\zb \to \zb^{-1}$ has the effect of exchanging $\delta_{f_1f_4} \delta_{f_2 f_3} \leftrightarrow \delta_{f_1f_3} \delta_{f_2 f_4}$ in \eqref{eq:G00def}, if one uses the  identity \eqref{eq:DhatInv}. 
As a consequence, the expansions \eqref{eq:G00tExp} and \eqref{eq:G00uExp} are identical up to this exchange of flavour indices.
}.
Since this expansion is identical to \eqref{eq:G00tExp} up to the exchange of flavour indices $\delta_{f_1f_4} \delta_{f_2 f_3} \leftrightarrow \delta_{f_1f_3} \delta_{f_2 f_4}$, one can simply repeat the procedure applied to the $t$-channel and obtain the full inversion function for \eqref{eq:G00def}
\begin{align} \label{eq:chhbSing}
	c(h,\hb) & =\frac{\delta_{f_1f_4} \delta_{f_2 f_3} +(-1)^k \delta_{f_1f_3} \delta_{f_2 f_4}}{4} \, \frac{\Gamma^2(h)}{2\Gamma(2h-1)} \,\times\nonumber \\*
	&\qquad\qquad \left[ \cI^{-1}(1-\hb) - \frac{1}{N} \Big(\cI^{-1}(1-\hb) + 2\mathcal{J}^{-1}(1-\hb) + \mathcal{J}^{-2}(1-\hb)\Big)\right]\,.
\end{align}

Functions of $\hb$ in \eqref{eq:chhbSing} contain poles that allow us to extract CFT data. 
Let us first consider 
\begin{align} \label{eq:zbInt}
	\cI^{a}(1-\hb) = \frac{\Gamma(2-2\hb)\,\Gamma^2(a+1)\, \Gamma(-\hb-a)}{\Gamma^2(1-\hb)\,\Gamma(2+a-\hb)}\,,
\end{align}
with generic $a$. 
The relevant poles%
\footnote{There are additional spurious poles due to $\Gamma(2- 2\hb)$ which we ignore \cite{Caron-Huot:2017vep, Simmons-Duffin:2017nub}.}
arise whenever the argument of  $\Gamma(-\hb-a)$ is a non-positive integer, where the gamma function behaves as
\begin{align}
	\Gamma(-\hb - a) \sim -\frac{(-1)^{\nb}}{\nb!} \, \frac{1}{\hb - (\nb-a)}\ , \qquad \hb \to \nb - a\ , \qquad \nb = 0,1,2, \ldots.
\end{align}
Expanding $\cI^a(1-\hb)$ around this value yields
\begin{align}
	\cI^{a}(1-\hb) \sim - \frac{ (-a)_{\nb}^{\,2}}{\nb! \left(\nb-2a -1\right)_{\nb}}\,\frac{1}{\hb - (\nb-a)}\,,
\end{align}
where $(x)_n = \Gamma(x+n)/{\Gamma(x)}  = (-1)^n {\Gamma(1-x)}/{\Gamma(1-x-n)}$ denotes the rising Pochhammer symbol.
Writing the behaviour of $\cI^a(1-\hb)$ near the pole in this form makes it manifest that the residue is a finite positive number if $a$ is a negative integer.
Using this result, we can analyse simple poles in \eqref{eq:chhbSing} at order $N^0$ and find 
\begin{align}
	c(h,\hb)\big|_{N^0} & \sim-\frac{1}{2}  \frac{\delta_{f_1f_4} \delta_{f_2 f_3} +(-1)^k \delta_{f_1f_3} \delta_{f_2 f_4}}{4} \,\frac{(\nb!)^2}{ (2\nb)!}\,
	\frac{\left[\left(\nb+ k\right)!\right]^2}{\left(2 \nb + 2k\right)!} \,\frac{1}{\hb - (1+\nb)}\,,
\end{align}
where we have used that $h$ and $\hb$ are related by $h = \hb + k$, from which it follows that as $\hb \to 1+ \nb$ so too $h \to 1+\nb + k$. After being inserted into \eqref{eq:cIh}, this results in the appearance $k$-dependent factorials.
By comparing the above expression with \eqref{eq:hhbPoles2}, we can extract the squares of generalised free field OPE coefficients
\begin{align} \label{eq:Cnk}
    \Big|c_{(0)(0,0)}^{\nb,k}\Big|^2 &=\frac{\delta_{f_1f_4} \delta_{f_2 f_3} +(-1)^k \delta_{f_1f_3} \delta_{f_2 f_4}}{4} \,\frac{(\nb!)^2}{ (2\nb)!}\, \frac{\left[\left(\nb+ k\right)!\right]^2}{\left(2 \nb + 2k\right)!}\,.
\end{align}
By setting $\nb =0$ one obtains
\begin{align} \label{eq:c0}
	\Big|c_{(0)(0,0)}^k\Big|^2 &= \frac{\delta_{f_1f_4} \delta_{f_2 f_3} +(-1)^k \delta_{f_1f_3} \delta_{f_2 f_4}}{4}\frac{(k!)^2}{(2k)!} \,,
\end{align}
which are the leading OPE coefficients for the exchange of minimal-twist operators and after using appropriate flavour projections \eqref{eq:projectors} we obtain \eqref{eq:c0ksingsing} and \eqref{eq:c0ksingsym}.

At order $1/N$, we expect \eqref{eq:chhbSing} to contain simple and double poles as $\hb \to 1 + \nb$.
In general, the location and the degree of such divergences in the inversion formula is determined by the behaviour of integrands in \eqref{eq:IntDef} as $Z, \Zb \to 0$, with double poles arising  due to the presence of $\log \Zb$ terms.
To see this, we can manipulate the logarithm function in \eqref{eq:IntDef2}  and obtain
\begin{align}
	\cJ^a(1-\hb) & = - \frac{d}{d a} \cI^a(1-\hb) + \sum_{m=1}^\infty\frac{(-1)^m}{m}\, \cI^{a-m}(1-\hb)\,.
	\label{eq:JaIntDec}
\end{align}
One finds that 
\begin{align} \label{eq:ddaI}
	- \frac{d}{d a} \cI^a(1-\hb) & = \cI^a(1-\hb)
	\Big[\psi(-a- \hb) + \psi(2 + a- \hb) - 2 \psi(a+1)\Big]\,,
\end{align}
where $\psi(z) \equiv \Gamma'(z)/\Gamma(z)$ is the digamma function, which diverges whenever its argument is a non-positive integer.
As such, the relevant double poles arise due to the combination of simple poles in both $\psi(-\hb-a)$ and $\Gamma(-\hb-a)$ in $\cI^a(1-\hb)$, with 
\begin{align}
	\label{eq:DoublePole}
	\Gamma(-a-\hb)\psi(-a - \hb)\sim - \frac{(-1)^{\nb}}{\nb!}\, \frac{1}{\left(\hb- (\nb-a)\right)^2} + \ldots\ , \qquad \nb = 0,1,2,\ldots
\end{align}
where the dots denote regular terms. 
In contrast, simple poles in $\cJ^a$ arise from three different places: from the series term in the second line of \eqref{eq:JaIntDec},  from $\Gamma(-\hb-a)$ in $\cI^a(1-\hb)$ when it is multiplied by the remaining two digamma functions in \eqref{eq:ddaI}, and from the first order terms  in the $\hb - (\nb-a)$ expansion of the prefactors multiplying \eqref{eq:DoublePole}
\begin{align}
	&\frac{\Gamma(2- 2\hb)\,\Gamma^2(1+a)}{\Gamma^2(1- \hb)\, \Gamma(2+ a- \hb)}\xrightarrow{\hb \to \nb - a}
	\frac{(-1)^{\nb}(-a)_{\nb}^{\,2}}{(\nb-2a -1)_{\nb}}\bigg[ 1 + \left( \hb - (\nb-a)\right) \times\nonumber\\
	& \big(2 \psi(1 + a - \nb) - 2 \psi(2 + 2a - 2\nb) + \psi(2 + 2a - \nb) \big)+ O\left(\left( \hb - (\nb-a)\right)^2\right)\bigg]\,.
	\label{eq:dp1}
\end{align}
After combining all these contributions, one finds that the pole structure of \eqref{eq:IntDef2} near $\hb = \nb - a$ is given by
\begin{equation}
	\label{eq:cJapoles}
	\cJ^a(1-\hb)  \sim -\frac{(-a)_{\nb}^{\,2}}{\nb! \,(\nb -2a -1)_{\nb}}\Biggr[\frac{1}{\left(\hb - (\nb-a)\right)^2}+ \frac{2 \psi(\nb - 2a -1) -2 \psi(2\nb -2a -1)}{\hb - (\nb -a)}\Biggr]\,.
\end{equation}
There is also an additional, spin-dependent, contribution coming from the prefactor \eqref{eq:cIh}
\begin{equation}
	\frac{\Gamma^2(h)}{\Gamma(2h-1)} \sim \frac{\Gamma^2(k + \nb - a)}{\Gamma(2 k + 2\nb - 2a -1)} \bigg[ 1 + 2\big(\hb - (\nb - a)\big)
	\Big( \psi(\nb +k-a) - \psi(2k +2\nb -2a -1)\Big)\!\bigg]\,,
\end{equation}
which is multiplying the $\cJ^a(1-\hb)$ in \eqref{eq:chhbSing}.
Thus all in all
\begin{align}
	&\frac{\Gamma^2(h)}{\Gamma(2h-1)} \cJ^{a}(1-\hb) \sim 
	-\frac{(-a)_{\nb}^{\,2}}{\nb! (\nb -2a -1)_{\nb}}\,
	\frac{\Gamma^2(k + \nb - a)}{\Gamma(2 k + 2\nb - 2a -1)} \Biggr[\frac{1}{\left(\hb - (\nb-a)\right)^2}\nonumber\\
	& + 2\frac{ \psi(\nb - 2a -1) -\psi(2\nb -2a -1)+ \psi(\nb +k-a) - \psi(2\nb + 2k  -2a -1) }{\hb - (\nb -a)}\Biggr]\,.
\end{align}
Applying this result to \eqref{eq:chhbSing} at order $1/N$, one finds 
\begin{align}\label{eq:Res1/N}
	&c(h, \hb)\Big|_{1/N} \sim\nonumber\\*
	& -\frac{1}{2} \frac{\delta_{f_1f_4} \delta_{f_2 f_3} +(-1)^k \delta_{f_1f_3} \delta_{f_2 f_4}}{4}\frac{(\nb!)^2}{ (2\nb)!}\,
	\frac{\left[\left(\nb+ k\right)!\right]^2}{\left(2 \nb + 2k\right)!}\Biggr\{ \frac{-\left(\nb^2+ \nb +2\right)}{\left(\hb - (\nb+1)\right)^2}- \frac{1}{\hb - (\nb +1)}\times \nonumber\\*
	& 
	\bigg[1+4\Big(\psi(\nb +1) -\psi(2\nb +1)+ \psi(\nb +k +1) - \psi(2\nb + 2k  +1)\Big)\\*
	& + 2\nb (\nb+1) \Big(\psi(\nb +2) -\psi(2\nb +1)+ \psi(\nb +k +1) - \psi(2\nb + 2k +1)\Big)\bigg]\Biggr\}+ \ldots\,,\nonumber
\end{align}
where the dots again denote regular terms.
By comparing with \eqref{eq:hhbPoles2}, one can read off 
\begin{subequations}
	\begin{align}
		\gamma_{(1),(0,0)}^{\nb,k} &= - \left(\nb^2+ \nb +2\right)\,,\label{eq:nkCFTDatag}\\
		\Big|c_{(1) (0,0)}^{\nb, k}\Big|^2 &= 
		- \Big|c_{(0)(0,0)}^{\nb,k}\Big|^2
		\bigg[1+4\Big(\psi(\nb +1) -\psi(2\nb +1)+ \psi(\nb +k +1) - \psi(2\nb + 2k  +1)\Big)\nonumber\\*
		& \qquad +2\nb (\nb+1) \Big(\psi(\nb +2) -\psi(2\nb +1)+ \psi(\nb +k +1) - \psi(2\nb + 2k  +1)\Big)\bigg]
		\,,\label{eq:nkCFTDatac}
	\end{align}
\end{subequations}
where we have used \eqref{eq:Cnk}. 
For $\nb=0$ this CFT data simplifies greatly, in particular, the anomalous dimensions are equal for all spins and are given by \mbox{$\gamma_{(1),(0,0)}^{k}= - 2$}, which for $k>2$ agrees with \eqref{eq:delksing} and \eqref{eq:c0ksingsym}. 
Similarly, the $1/N$ corrections to the coefficients in \eqref{eq:coeffexpc} can be written in terms of harmonic numbers as
\begin{align}
 	\Big|c_{(1)(0,0)}^{k}\Big|^2 = \Big|c_{(0)(0,0)}^{k}\Big|^2\Big(4H_{2k} -4H_k-1\Big) \,,
\end{align}
which, after suitable flavour projections, match \eqref{eq:c1ksingsing} and \eqref{eq:c1ksingsym} provided the spin is large enough.

One can also apply the above procedure to other R-symmetry projections of \eqref{eq:CCabfi}. 
For the $(1,0)$ projection, the anomalous dimensions and coefficients of global blocks to order $1/N$ in different flavour projections are given in equations \eqref{eq:c0ktripsing}--\eqref{eq:c1ktripsym}. 
The most important difference with respect to the R-symmetry singlet is an additional relative minus factor between the $t$-channel and $u$-channel contributions to the inversion function, which correctly reproduces non-zero results only for exchanges of operators with odd spin. 
On the other hand, minimal twist operators exchanged in $(0,1)$ and $(1,1)$ R-symmetry projections of \eqref{eq:CCabfi} are protected by supersymmetry and correspondingly the inversion formula shows that anomalous dimensions are non-zero only for $\nb\geq 1$ and that for $\nb=0$ the OPE coefficients match free field results.

\bibliographystyle{utphys}
\bibliography{2105.04670v2.bib}
\end{document}